\documentclass[12pt]{article}

\usepackage{graphicx}
\usepackage[pdf]{pstricks}
\usepackage{amsmath,amssymb,amsfonts,graphicx}
\usepackage{ucs}
\usepackage{soul}
\usepackage{epstopdf}
\usepackage{color}
\usepackage[utf8]{inputenc}
\usepackage{amsthm}
\usepackage{fontenc}
\usepackage{tikz}
\RequirePackage{slashed}
\usepackage{cancel}
\usepackage[normalem]{ulem}

\usepackage{color}
\usepackage{latexsym}
\usepackage{amssymb}
\usepackage{booktabs}
\usepackage{comment}
\usepackage{authblk}
\usepackage{float} 
\usepackage{cite}%regroups citations

\usepackage[a4paper,top=2cm,bottom=2cm,left=2cm,right=2cm]{geometry}

\usepackage{braket}

\newcommand{\bit}{\begin{itemize}}
\newcommand{\eit}{\end{itemize}}

\newlength\savedwidth

\title{Double parton distributions in the pion 
\\
in the Nambu and Jona-Lasinio model
}

\begin{document}

\author[1]{Aurore Courtoy}
\affil[1]{\small{ Instituto de F\'isica, Universidad Nacional Aut\'onoma de M\'exico\\
Apartado Postal 20-364, 01000 Ciudad de M\'exico, Mexico} }
\author[2]{Santiago Noguera}
\affil[2]{\small{ Departamento de F\`isica Te\`orica and IFIC, Centro Mixto Universidad de Valencia-CSIC,
E-46100 Burjassot (Valencia), Spain } }
\author[3]{Sergio Scopetta}
\affil[3]{ \small{Dipartimento di Fisica e Geologia, Universit\`a degli Studi di Perugia, and Istituto Nazionale 
di Fisica 
Nucleare,
Sezione di Perugia, Via A. Pascoli, I-06123, Perugia, Italy } }

\maketitle

\begin{abstract}
{
Two-parton correlations { in the pion},
a non perturbative information encoded 
in double parton 
distribution functions,
are
investigated in the Nambu and Jona-Lasinio model.  
It is found that double parton distribution functions 
expose novel dynamical information 
on the structure of the pion,   not accessible 
through one-body parton distributions, as it happens
in several estimates for the proton target
and in a previous evaluation for the pion, in a light-cone framework.
 Expressions and predictions are given for double parton distributions corresponding
to leading-twist Dirac operators in the quark vertices, and to
different regularization methods for the Nambu and Jona-Lasinio model. { These results are particularly relevant in view of
forthcoming lattice data.}}

\end{abstract}

\section{\label{sec:intro}Introduction}

Since the start of the LHC operation,
multiple parton interactions (MPI) have become an important 
topic in nowadays hadronic Physics \cite{Bartalini:2017jkk}.
Due to the high partonic densities reached, processes where
more than two partons from the two colliding protons
participate in the actual scattering process are likely to happen. 
The simplest form of MPI, 
double parton scattering (DPS), 
involving two simultaneous hard collisions,
has been indeed  observed at the LHC (see, e.g., Ref. \cite{Aad:2013bjm}).
The DPS cross section is written in terms of double 
parton distribution functions (dPDFs) \cite{Paver:1982yp,Diehl1},
related to the number density of two partons,
with given longitudinal momentum fractions,
located at a given 
transverse separation in 
coordinate space. 
These distributions encode information complementary to that
obtained through the tomography, accessed using
electromagnetic probes, 
in terms of generalized parton distributions (GPDs)
\cite{Guidal:2013rya,Dupre:2016mai}.  
If measured, dPDFs would therefore represent 
a novel tool to study the three-dimensional hadron structure \cite{blok1,blok2,fabbro,ffe}.
Indeed, they are 
{{sensitive}} 
to  two-parton correlations {{not accessible via one body 
distributions, i.e.  standard  partons distribution functions (PDFs) 
and GPDs} } (see Ref. \cite{Kasemets:2017vyh}
for a recent report).
Since dPDFs describe soft
Physics, they are non perturbative objects and have
not been evaluated in 
QCD. It is therefore useful to
estimate them at low momentum scales ($\sim \Lambda_{QCD}$),
using models of the hadron structure, as it has been proposed,
for the proton, in Refs. 
\cite{bag,noi2,noi1,kase,plb,JHEP2016,Traini:2016jru}. 
In order to match theoretical predictions with future experimental 
analyses, 
the results of these 
calculations are then evolved using perturbative QCD to reach the 
high momentum scale 
of the data. Evolution properties
of dPDFs have been studied {in the past}  \cite{Kirschner:1979im,Shelest:1982dg} and have been recently object
of deep investigation (for new developments see, e.g., the report
\cite{Diehl:2017wew}).

%Due to the difficulty of measuring pion observables,the interest for pion dPDFs is mainly theoretical.  
 
 Recently, results have been
obtained for two current correlations in the pion on the lattice
\cite{Zimmermann:2017ctb,Bali:2018nde}, quantities
related to dPDFs,
whereas 
the corresponding evaluation for the nucleon appears much more involved.
The novel possibility to compare results
with lattice data makes model calculations of pion dPDFs of relevant
theoretical interest.
A first estimate of pion dPDFs has been performed in Ref.
\cite{Rinaldi:2018zng},
using light cone wave functions obtained within the AdS/QCD correspondence.

In this paper we analyze pion dPDFs 
within a Nambu--Jona-Lasinio (NJL) framework
\cite{Klevansky:1992qe}.

 The dependence of the results on the possible choice of regularization
of the NJL model is investigated within two different prescriptions,
the standard Pauli--Villars one and a properly built
Light Front regularization.
Our aim is the first evaluation of pion dPDFs in a field theoretical approach which
allows to study systematically different contributions. This procedure is useful, e.g., to
check the validity of approximated expressions for dPDFs
in terms of GPDs,
and to calculate quantities related to two current correlations in the pion,
towards a direct comparison with lattice data.

The paper is structured as follows.
In section \ref{II} we 
define the pion dPDFs, we describe the NJL evaluation scheme and show the results of the calculation.
In section \ref{III} we test the validity,
within our scheme,
of a commonly used approximation
of the dPDF in terms of GPDs.
Conclusions are collected in section \ref{IV}.

\section{\label{II}
Double parton distribution functions in the pion}

For the definition of the quantities to be evaluated, we follow the conventions introduced in Ref. \cite{Diehl1}.
In particular, a dPDF for the pion is defined as

\begin{eqnarray}
F_{a_1 ,\bar a_2}(x_1,x_2,\vec y_\perp) = \int d^2 k_{1 \perp} d^2 k_{2 \perp}
F_{a_1, \bar a_2}(x_1,x_2,\vec k_{1 \perp}, \vec k_{2 \perp},\vec y_{\perp})~,
\label{uno}
\end{eqnarray}
with $\vec y_\perp$ the transverse distance between 
the two partons,
starting from the generic light-cone correlator

\begin{eqnarray}
F_{a_1, \bar a_2}(x_1,x_2,\vec k_{1 \perp}, \vec k_{2 \perp},\vec y_\perp) & = & - 2 P^+
\int \frac{d z_1^- d^2 z_{1 \perp}}{ (2 \pi)^3 }
\frac{d z_2^- d^2 z_{2 \perp}}{ (2 \pi)^3 }
e^{i x_1 z_1^- P^+ -i \vec z_{1 \perp}\cdot
\vec k_{1 \perp}}
\nonumber
\\
& \times &
e^{i x_2 z_2^- P^+ -i \vec z_{2 \perp}\cdot
\vec k_{2 \perp}} \, {\cal{O}}_{a_1 \bar a_2} ~,
\end{eqnarray}
where

\begin{eqnarray}
{\cal{O}}_{a_1 , \bar a_2}
& = & \int d y^-
\,\left\langle \pi^{i}\left(P\right)\right|
\left \{ \left [ \bar{q}\left(\tfrac{1}{2}z_{2}\right)\bar{\Gamma}_{\bar{a}_{2}}q\left(-\tfrac{1}{2}z_{2}\right)\right]
\right .
\nonumber
\\
& \times &
\left .
\left [ \bar{q}\left(y-\tfrac{1}{2}z_{1}\right)\bar{\Gamma}_{a_{1}}q\left(y+\tfrac{1}{2}z_{1}\right)\right]\right \} _{z_{1}^{+}=z_{2}^{+}
=y^{+}=0}\left|\pi^{i}\left(P\right)\right\rangle \, ,
\label{due}
\end{eqnarray}
with $i$ the isospin index of the pion. In the equations above,
the index $a_{1}\,\left(\bar{a}_{2}\right)$
is a short notation for the Dirac and isospin  indices associated to
the matrices involved in the quark (antiquark) vertex:
\begin{equation}
\bar{\Gamma}_{a_{1}}=\Gamma_{a_{1}}\,\tau^{a_{1}} \, ,
\label{gammatau}
\end{equation}
with $\Gamma_{a_{1}}$ given by

\begin{eqnarray}
\Gamma_q & = & \Gamma_{\bar q} = \frac{1}{2} \gamma^+~,
\quad \quad \quad \quad 
\Gamma_{\Delta_q} = - \Gamma_{\Delta \bar q} = \frac{1}{2} \gamma^+ \gamma_5~,
\nonumber
\\
\Gamma_{\delta_q}^i & = & \Gamma_{\delta \bar q}^i = \frac{1}{2} i \sigma^{i+} \gamma_5~,
\label{tre}
\end{eqnarray}
in the vector, axial and tensor sectors,  related
to leading twist distributions, respectively.

 We follow Ref.~\cite{Diehl1} to analyze
the structure of the dPDFs. 
The functions 
$F_{q,\bar{q}}\left(x_{1},x_{2},y_{\perp}\right)$ and
$F_{\Delta q,\Delta\bar{q}}\left(x_{1},x_{2},y_{\perp}\right)$
are, by construction, scalar quantities and, therefore, { functions of $y_{\perp}=\left|\vec{y}_{\perp}\right|$.}
The two functions $F_{q,\Delta\bar{q}}\left(x_{1},x_{2},y_{\perp}\right)$
and $F_{\Delta q,\bar{q}}\left(x_{1},x_{2},y_{\perp}\right)$ vanish identically,
for parity conservation.
Four dPDFs  turn out to be two-dimensional vectors in the transverse plane.
They can be written in terms of scalar functions, as follows
\begin{equation}
\begin{array}{ccc}
F_{q,\delta\bar{q}^{j}}\left(x_{1},x_{2},\vec{y}_{\perp}\right)=\epsilon^{j,k}\,\hat{y}^{k}_{\perp}\,F_{q,\delta\bar{q}}^{v}\left(x_{1},x_{2},y_{\perp}\right)\,, &  & F_{\delta q^{j},\bar{q}}\left(x_{1},x_{2},\vec{y}_{\perp}\right)=\epsilon^{j,k}\,\hat{y}^{k}_{\perp}\,F_{\delta q,\bar{q}}^{v}\left(x_{1},x_{2},y_{\perp}\right)\quad,\\
\\
F_{\Delta q,\delta\bar{q}^{j}}\left(x_{1},x_{2},\vec{y}_{\perp}\right)=\hat{y}^{j}_{\perp}\,F_{\Delta q,\delta\bar{q}}^{v}\left(x_{1},x_{2},y_{\perp}\right)\,, &  & F_{\delta q^{j},\Delta\bar{q}}\left(x_{1},x_{2},\vec{y}_{\perp}\right)=\hat{y}^{j}_{\perp}\,F_{\delta q,\Delta\bar{q}}^{v}\left(x_{1},x_{2},y_{\perp}\right)\quad.
\label{VectorStr_y}
\end{array}
\end{equation}
The last dPDF is a tensor quantity and, in terms of scalar functions, it reads
\begin{equation}
F_{\delta q^{j},\delta\bar{q}^{k}}\left(x_{1},x_{2},\vec{y}_{\perp}\right)=\delta^{j,k}\,F_{\delta q,\delta\bar{q}}^{s}\left(x_{1},x_{2},y_{\perp}\right)+\left(2\hat{y}^{j}_{\perp}\,\hat{y}^{k}_{\perp}-\delta^{j,k}\right)\,F_{\delta q,\delta\bar{q}}^{t}\left(x_{1},x_{2},y_{\perp}\right)\quad, \label{TensorStr_y}
\end{equation}
with $\hat{y}_{\perp}=\vec{y}_{\perp}/\left|\vec{y}_{\perp}\right|$.

It is convenient to calculate the dPDFs
in momentum space,

\begin{eqnarray}
F_{a_1, \bar a_2}(x_1,x_2,\vec q_\perp) = \int d^2 k_{1 \perp} d^2 k_{2 \perp} 
d^2 y_{\perp}
e^{i\, \vec y_\perp \cdot \vec q _ \perp}
F_{a_1 ,\bar a_2}(x_1,x_2,\vec k_{1 \perp}, \vec k_{2 \perp},\vec y_{\perp})~,
\label{quattro}
\end{eqnarray}
a quantity often called ``2GPD" \cite{blok1,blok2}  which, at variance
with the dPDF Eq.~(\ref{uno}), is not related to a 
probability density. The natural support  in $x_{1,2}$  of the function $F_{q_1 \bar q_2}$  is 
$0 \leq x_{1,2} \leq 1$, $0 \leq x_1+x_2 \leq 1$.
The quantity $\vec{q}_\perp$ represents the imbalance
between the relative momentum of the two partons 
in the considered hadronic state and in its conjugated
one.

Expressions equivalent to Eqs.~(\ref{VectorStr_y}) 
and~(\ref{TensorStr_y}), yielding vector and tensor quantities
in terms of scalar functions, can be given in momentum
space as follows
\[
\begin{array}{ccc}
F_{q,\delta\bar{q}^{j}}\left(x_{1},x_{2},\vec{q}_{\perp}\right)=i\,\epsilon^{j,k}\,\hat{q}^{k}_{\perp}\,F_{q,\delta\bar{q}}^{v}\left(x_{1},x_{2},q_{\perp}\right)\,, &  & F_{\delta q^{j},\bar{q}}\left(x_{1},x_{2},\vec{q}_{\perp}\right)=i\,\epsilon^{j,k}\,\hat{q}^{k}_{\perp}\,F_{\delta q,\bar{q}}^{v}\left(x_{1},x_{2},q_{\perp}\right)\quad,\\
\\
F_{\Delta q,\delta\bar{q}^{j}}\left(x_{1},x_{2},\vec{q}_{\perp}\right)=i\,\hat{q}^{j}_{\perp}\,F_{\Delta q,\delta\bar{q}}^{v}\left(x_{1},x_{2},q_{\perp}\right)\quad, &  & F_{\delta q^{j},\Delta\bar{q}}\left(x_{1},x_{2},\vec{q}_{\perp}\right)=i\,\hat{q}^{j}_{\perp}\,F_{\delta q,\Delta\bar{q}}^{v}\left(x_{1},x_{2},q_{\perp}\right)\quad,
\end{array}
\]
\begin{equation}
F_{\delta q^{j},\delta\bar{q}^{k}}\left(x_{1},x_{2},\vec{q}_{\perp}\right)=\delta^{j,k}\,F_{\delta q,\delta\bar{q}}^{s}\left(x_{1},x_{2},q_{\perp}\right)+\left(2\hat{q}^{j}_{\perp}\,\hat{q}^{k}_{\perp}-\delta^{j,k}\right)\,F_{\delta q,\delta\bar{q}}^{t}\left(x_{1},x_{2},q_{\perp}\right)\,\,\,.\label{TensorStr_q}
\end{equation}

By evaluating
the Fourier transformations, Eq.~(\ref{quattro})
for the scalar, vector and tensor quantities,
the scalar functions
$F_{a_{1},\bar{a}_{2}}\left(x_{1},x_{2},y_{\perp}\right)$ in coordinate
space are found to be related to the scalar functions 
$F_{a_{1},\bar{a}_{2}}\left(x_{1},x_{2},q_{\perp}\right)$
in momentum space according to expressions given here below.
In the case of the scalar quantities $F_{q,\bar{q}}\left(x_{1},x_{2},q_{\perp}\right)$, $F_{\Delta q,\Delta\bar{q}}\left(x_{1},x_{2},q_{\perp}\right)$ 
 and 
%$F_{q,\Delta\bar{q}}\left(x_{1},x_{2},q_{\perp}\right)$,
%$F_{\Delta q,\bar{q}}\left(x_{1},x_{2},q_{\perp}\right),$\textcolor{red}{{}
%}
$F_{\delta q,\delta\bar{q}}^{s}\left(x_{1},x_{2},q_{\perp}\right),$
one gets
\begin{equation}
F_{a_{1},\bar{a}_{2}}\left(x_{1},x_{2},y_{\perp}\right)=\int\,\frac{dq_{\perp}\,q_{\perp}}{2\pi}J_{0}\left(q_{\perp}y_{\perp}\right)\,F_{a_{1},\bar{a}_{2}}\left(x_{1},x_{2},q_{\perp}\right)\,\,.
\label{fts}
\end{equation}
For the scalar functions defining the vector quantities, $F_{q,\delta\bar{q}}^{v}\left(x_{1},x_{2},q_{\perp}\right)$,
$F_{\delta q,\bar{q}}^{v}\left(x_{1},x_{2},q_{\perp}\right)$
\\
$F_{\Delta q,\delta\bar{q}}^{v}\left(x_{1},x_{2},q_{\perp}\right)$
and $F_{\delta q,\Delta\bar{q}}^{v}\left(x_{1},x_{2},q_{\perp}\right)$,
one has
\begin{equation}
F_{a_{1},\bar{a}_{2}}^{v}\left(x_{1},x_{2},y_{\perp}\right)=\int\,\frac{dq_{\perp}\,q_{\perp}}{2\pi}J_{1}\left(q_{\perp}y_{\perp}\right)F_{a_{1},\bar{a}_{2}}^{v}\left(x_{1},x_{2},q_{\perp}\right)\,\,,
\label{ftv}
\end{equation}
and, for the scalar function present in the tensor structure, $F_{\delta q,\delta\bar{q}}^{t}\left(x_{1},x_{2},q_{\perp}\right)$,
the relation is
\begin{equation}
F_{\delta q,\delta\bar{q}}^{t}\left(x_{1},x_{2},y_{\perp}\right)=-\int\,\frac{dq_{\perp}\,q_{\perp}}{2\pi}J_{2}\left(q_{\perp}y_{\perp}\right)F_{\delta q,\delta\bar{q}}^{t}\left(x_{1},x_{2},q_{\perp}\right)\,\,.
\label{ftt}
\end{equation}

The calculation framework is the NJL
model, the most realistic model for the pseudoscalar
mesons based on a local
quantum field theory built with quarks \cite{Klevansky:1992qe}. It respects the 
realization of chiral symmetry and
gives a good description of low energy properties. Mesons are described as bound states, in
a fully covariant fashion, using the Bethe-Salpeter amplitude,
in a field theoretical framework.
In this way, Lorentz covariance is preserved.  The NJL model is a
non-renormalizable field theory and therefore a {regularization procedure has to be implemented.}
%cut-off procedure has to be implemented.
%\textcolor{red}{\sout{Here, the Pauli-Villars regularization scheme, which respects the %gauge
%symmetry of the problem, has been adopted} 
 We have performed initially our calculations in the Pauli- Villars ($PV$) regularization
scheme, which is a well established method.
%We have also built a regularization method for the NJL mode adapted for light front %calculations. For comparison,
%we will give results in both methods..
The NJL model, together with its 
regularization procedure, can
be regarded as an effective theory of QCD.
Some basic features of the NJL model and details on the regularization schemes are 
reported in Appendix A. 

Model calculations of meson partonic structure within
this approach have a long story of successful predictions
\cite{Davidson:2001cc,Theussl:2002xp,
RuizArriola:2002bp,Courtoy:2007vy,
Courtoy:2008nf,Courtoythesis,
Noguera:2011fv,Weigel:1999pc,
ns,
Broniowski:2017gfp,Ceccopieri:2018nop}.

One should remember that collinear parton distributions 
obtained within a model have to be associated to a low 
momentum scale $Q_0^2$, at which one has only valence quarks, and, 
in order to be used to predict measured quantities,
have to be evolved to higher momentum scales according to
perturbative QCD (pQCD).

Let us describe the main steps of the calculation of
Eq.~\eqref{quattro} in the NJL model.
For the pion $i$ we use the state
\begin{equation}
\left|\pi^{i}\left(P\right)\right\rangle =\int d^{4}y_{1}\,d^{4}y_{2}\,\frac{d^{4}k}{\left(2\pi\right)^{4}}e^{-i\frac{1}{2}P\cdot\left(y_{1}+y_{2}\right)}\,e^{-ik\cdot\left(y_{1}-y_{2}\right)}\bar{q}\left(y_{1}\right)\phi_{\pi^{i}}\left(k,P\right)q\left(y_{2}\right)\left|0\right\rangle \, ,
\label{A.5}
\end{equation}
with $\phi_{\pi^{i}}$ the quark-pion vertex function for a $\pi^i$.
In the NJL model 
the amplitude
$\phi_{\pi^i}\left(k,P\right)$ is independent on the relative
and total quark-antiquark momenta, $k$
and $P,$ respectively, and we have 
\begin{equation}
\phi_{\pi^{i}}\left(k,P\right)=ig_{\pi qq}i\gamma_{5}\tau^{i}
\label{ampli}
\end{equation}
where $g_{\pi qq}$ is the quark-pion coupling constant and $\tau^{i}$
is the isospin matrix associated to the corresponding pion $\pi^i$.

\begin{figure}
\centering{}\includegraphics[scale=0.5]{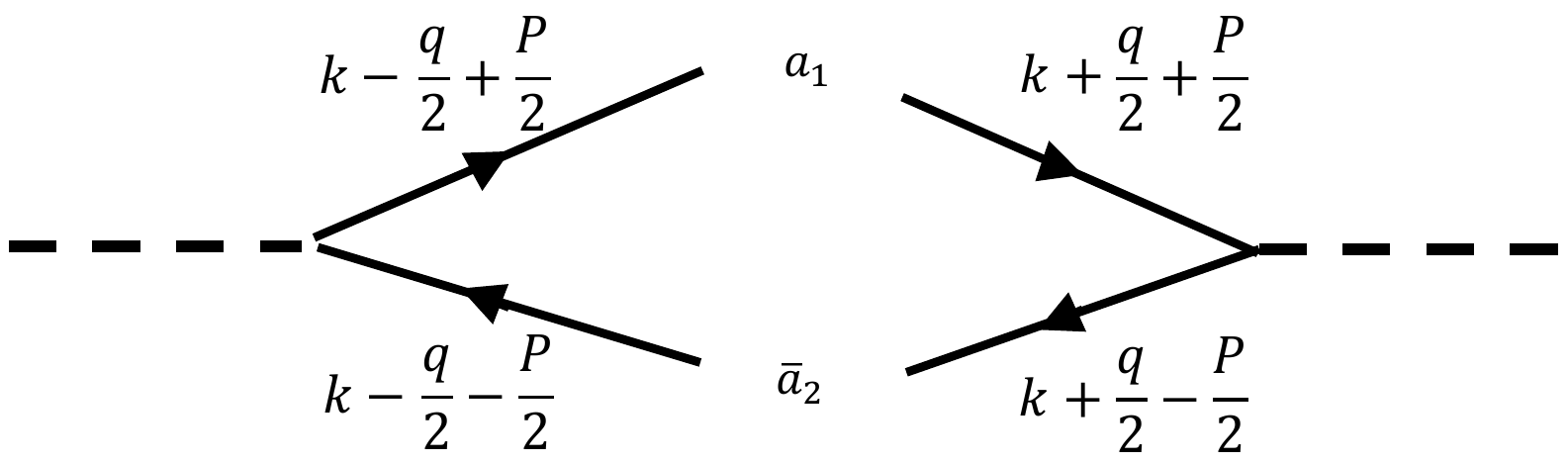}\caption{Diagram of the Double Parton Distribution Function associated to Eq.~(\ref{A.7}). Open vertices represent nonlocal current insertions. }
\label{Fig_diagrama_rombo} 
\end{figure}

Let us consider the case of a $\pi^{+}$, and
therefore the operators $\bar{\Gamma}_{a}
= \Gamma_a \frac{1}{2} (1 + \tau_3)$ and
$\bar{\Gamma}_{\bar a}
= \Gamma_{\bar a} \frac{1}{2} (1 - \tau_3)$.
At leading order, we have the contribution
depicted in Fig \ref{Fig_diagrama_rombo}. 
Using Eqs.~\eqref{A.5} and~\eqref{ampli} in Eq.~\eqref{due},
after a tedious but straightforward
calculation we get, for the momentum space
dPDF Eq.~\eqref{quattro},
{
\begin{align}
F_{a_1,\bar{a_2}}\left(x_{1},x_{2},\vec{q}_{\perp}\right) & =-2P^{+}\int\frac{d^{4}k}{\left(2\pi\right)^{4}}\frac{dq^{-}}{2\pi}\delta\left(P^{+}x_{1}-\frac{1}{2}\left(P+2k\right)^{+}\right)\,\delta\left(P^{+}x_{2}+\frac{1}{2}\left(-P+2k\right)^{+}\right)\nonumber \\
 & \left(-\right)\text{Tr}\left[iS_{F}\left(\frac{P}{2}+k-\frac{q}{2}\right)\,\phi_{\pi^{+}}\,iS_{F}\left(-\frac{P}{2}+k-\frac{q}{2}\right)\,\bar{\Gamma}_{\bar{a_2}}\,iS_{F}\left(-\frac{P}{2}+k+\frac{q}{2}\right)\right.
 \nonumber \\ & \times
 \left .
 \bar{\phi}_{\pi^{+}}\, iS_{F}\left(\frac{P}{2}+k+\frac{q}{2}\right)\bar{\Gamma}_{a_1}\right]
 \, ,\label{A.7}
\end{align}
}
where
\begin{equation}
\bar{\phi}_{\pi^{+}}\left(k,P\right)=-\gamma_{0}\,\phi_{\pi^{+}}^{\dagger}\left(k,P\right)\,\gamma_{0}~,
\end{equation}
 and the quark propagator is given by $S_F(k) =(\cancel{k}-m
+ i \epsilon)^{-1}$  and the trace is intended in color, isospin and quadri-spinor indices. 

We observe that this contribution can be obtained defining 
\begin{equation}
F_{a_1,\bar{a_2}}\left(x_{1},x_{2},\vec{q}_{\perp}\right)=-2P^{+}\int\frac{dq^{-}}{2\pi}\,\mathcal{T} \, ,\label{A.9}
\end{equation}
where $\mathcal{T}$ is the Feynman amplitude corresponding to the
diagram of Fig.~\ref{Fig_diagrama_rombo}, $q^{\mu}=\left(0,\vec{q}_{\perp},q^{-}\right)$
and the following bilocal current vertex
has to be used:

\begin{table}[htbp]
\begin{centering}
\begin{tabular}{ccc}
\includegraphics[scale=0.5]{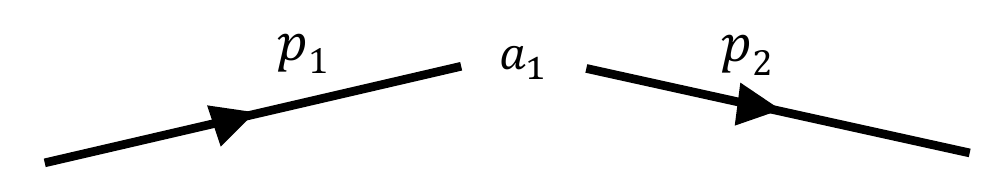}  & $\Rightarrow$  & $\bar{\Gamma}_{a_{1}}\,\delta\left(P^{+}x_{1}-\frac{1}{2}\left(p_{1}+p_{2}\right)^{+}\right)$\tabularnewline
\end{tabular}
\par\end{centering}
\begin{centering}
\begin{tabular}{ccc}
\includegraphics[scale=0.5]
{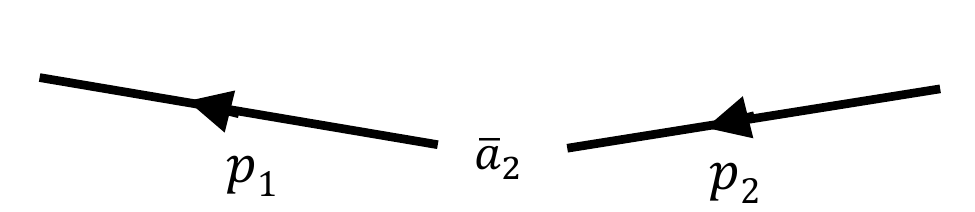}  & $\Rightarrow$  & $\bar{\Gamma}_{\bar{a}_{2}}\,\delta\left(P^{+}x_{2}+\frac{1}{2}\left(p_{1}+p_{2}\right)^{+}\right)$\tabularnewline
\end{tabular}
\par\end{centering}
\caption{Bare vertices associated to the bilocal currents.}
\label{Tab_bare_vertex} 
\end{table}

The integration over the $k^{+}$ variable in Eq.~(\ref{A.7})
using the Dirac delta, and the integration over $k^{-}$ and $q^{-}$
using the Cauchy theorem of residues, give
%\textcolor{red}{[a partir de aquí cambia tanto que no dejo lo antiguo]}
\begin{align}
& F_{a_{1},\bar{a_{2}}}\left(x_{1},x_{2},\vec{q}_{\perp}\right)  =\delta\left(x_{1}+x_{2}-1\right)\,\theta\left(x_{1}\right)\,\theta\left(1-x_{1}\right)\left(-\frac{N_{c}\,g_{\pi qq}^{2}}{2\pi P^{+2}}\right) \label{A.10} \\
& \int\frac{d^{2}k_{\perp}}{\left(2\pi\right)^{2}}  \frac{tr_{a_{1},\bar{a}_{2}}\left(\vec{k}_{\perp},\vec{q}_{\perp}\right)}{\left[\left(\vec{k}_{\perp}-\frac{q_{\perp}}{2}\right)^{2}+m^{2}-x_{1}\left(1-x_{1}\right)m_{\pi}^{2}-i\epsilon\right]\left[\left(\vec{k}_{\perp}+\frac{q_{\perp}}{2}\right)^{2}+m^{2}-x_{1}\left(1-x_{1}\right)m_{\pi}^{2}-i\epsilon\right]}\,,\nonumber
\end{align}
with 
\begin{eqnarray}
tr_{a_1,\bar{a_2}}\left(\vec{k}_{\perp},\vec{q}_{\perp}\right) & = & tr\left[\left(\frac{\cancel{P}}{2}+\cancel{k}-\frac{\cancel{q}}{2}+m\right)\gamma_{5}\left(-\frac{\cancel{P}}{2}+\cancel{k}-\frac{\cancel{q}}{2}+m\right)\Gamma_{\bar{a}_{2}}
\right. \nonumber \\
& \times & \left .
\left(-\frac{\cancel{P}}{2}+\cancel{k}+\frac{\cancel{q}}{2}+m\right)\gamma_{5}\left(\frac{\cancel{P}}{2}+\cancel{k}+\frac{\cancel{q}}{2}+m\right)\Gamma_{a_{1}}\right] \, ,
\label{A.11}
\end{eqnarray}
where $tr$ implies trace over quadri-spinors indices. Besides,
we have
\begin{equation}
k^{\mu}=\left(\left(x_{1}-\frac{1}{2}\right)P^{+},\vec{k}_{\perp},\frac{P^{-}}{2}-\frac{m^{2}+\vec{k}_{\perp}^{\,2}}{2P^{+}\left(1-x_{1}\right)} \right) \, ,
\end{equation}
\begin{align}
q^{\mu} & =\left(0,\,\vec{q}_{\perp},\,\frac{-\vec{k}_{\perp}\cdot\vec{q}_{\perp}}{P^{+}\left(1-x_{1}\right)} \right) \, ,
\end{align}
and $P^{2}=2P^{+}P^{-}=m_{\pi}^{2}.$

\begin{figure}
\begin{center}
\includegraphics[scale=0.5]{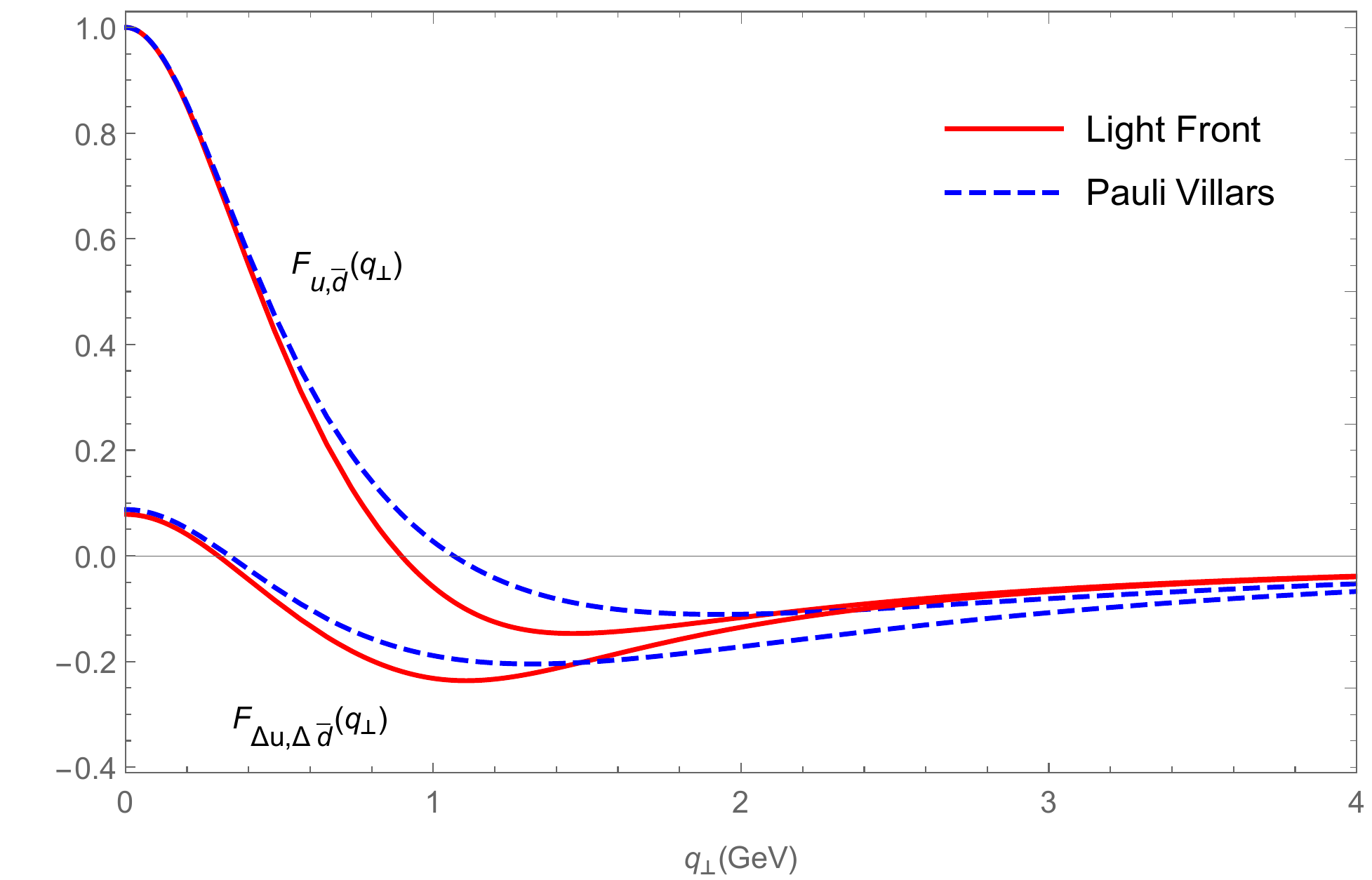}
\vskip 4.mm
\includegraphics[scale=0.5]{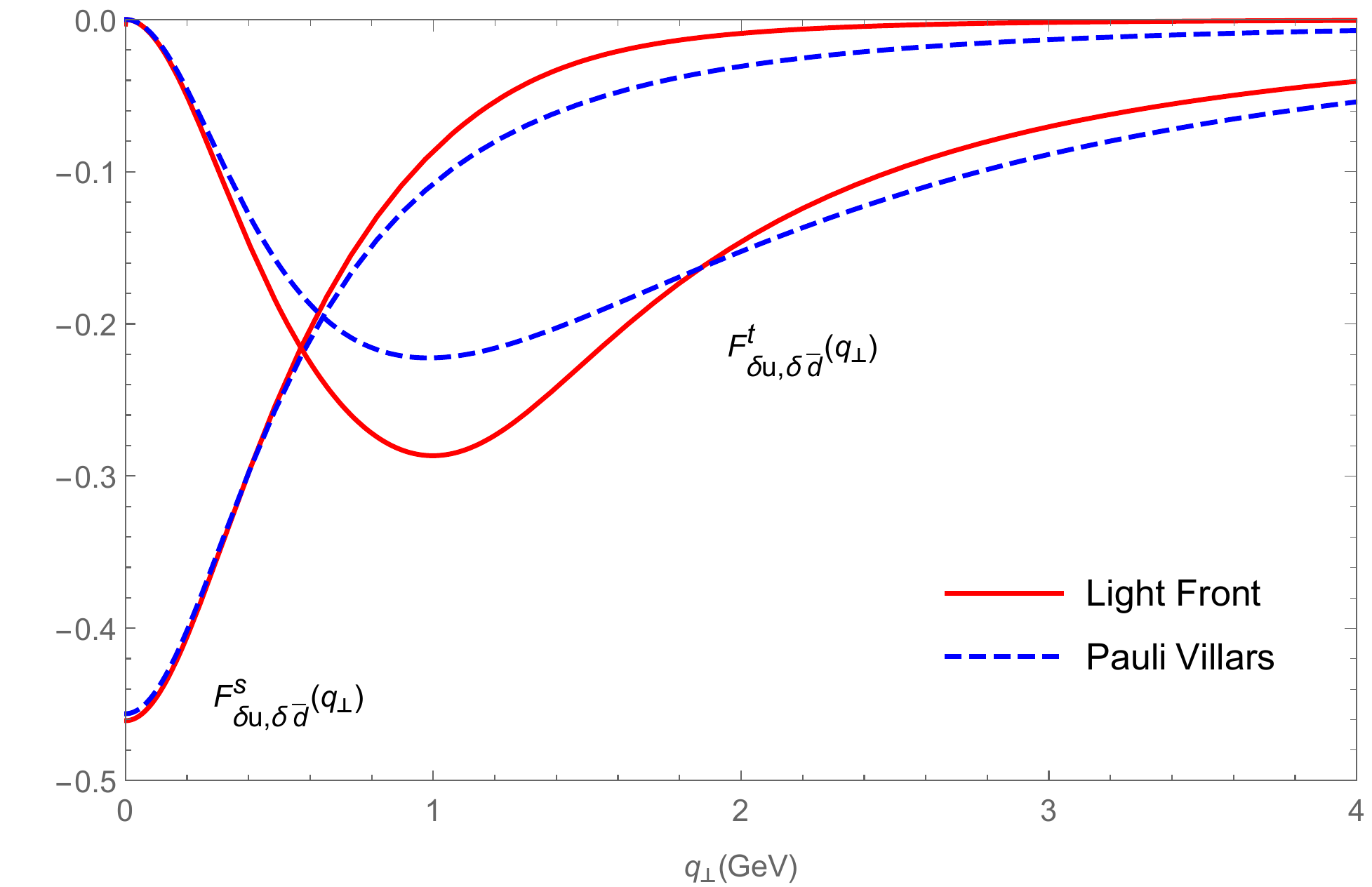}
\vskip 4.mm
\includegraphics[scale=0.5]{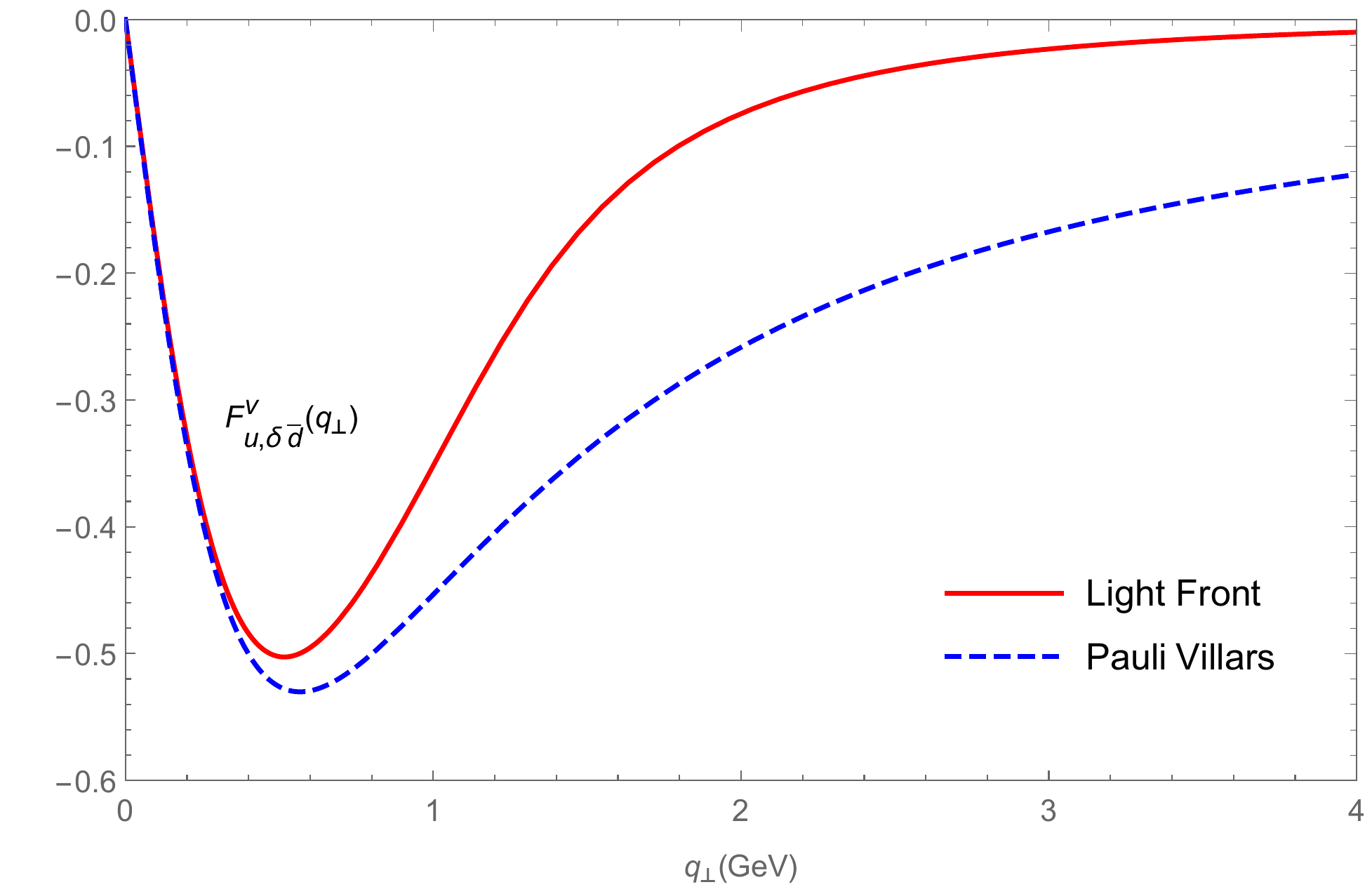}
\caption{Double parton distribution functions in the NJL model, 
in momentum space,
integrated over $x_1, x_2$.
{Dashed (full)} curves represent results obtained using the Pauli-Villars (Light-Front)
regularization method. Top panel: quantities obtained integrating
$F_{u,\bar{d}}\left(x_{1},x_{2},q_{\perp}\right)$, Eq. (\ref{A.16}) (Eq. (\ref{A.16-1}))
and
$F_{\Delta u,\Delta \bar{d}}\left(x_{1},x_{2},q_{\perp}
\right)$, Eq. (\ref{aa}) (Eq. (\ref{aa-1}));
central panel:
quantities obtained integrating
$F_{\delta u,\delta \bar{d}}^{s}\left(x_{1},x_{2},q_{\perp}
\right)$, Eq. (\ref{tt0}) (Eq. (\ref{tt0-1}))
and
$F_{\delta u,\delta \bar{d}}^{t}\left(x_{1},x_{2},q_{\perp}
\right)$, Eq. (\ref{tt2}) (Eq. (\ref{tt2-1}));
bottom panel: the
quantity obtained integrating
$F_{u,\delta \bar{d}}^{v}\left(x_{1},x_{2},q_{\perp}
\right)$, Eq. (\ref{vt}) (Eq. (\ref{vt-1})).}
\end{center}
\label{Fig_dPDF_q}
\end{figure}

The integral over $k_{\perp}$ present in Eq (\ref{A.10}) is then
rendered finite using the adopted Pauli-Villars regularization method,
described in the Appendix \ref{App.NJL_PV_regularization}.
Some intermediate steps are given in Appendix \ref{App.Intermediate-Results}. Our final results for the  scalar functions appearing in the dPDFs related to  different Dirac structures in the quark vertices are
\begin{equation}
F_{u,\bar{d}}^{\left(PV\right)}\left(x_{1},x_{2},q_{\perp}\right)=C(x_{1},x_{2})\,\sum_{j=0}^{2}c_{j}\,\left\{ -\ln\frac{\kappa_{j}}{\kappa}+\left[2\frac{m_{\pi}^{2}}{q_{\perp}^{2}}x_{1}(1-x_{1})-1\right]f\left(\frac{\kappa_{j}}{q_{\perp}^{2}}\right)\right\} \,,\label{A.16}
\end{equation}
\begin{equation}
F_{\Delta u,\Delta\bar{d}}^{\left(PV\right)}\left(x_{1},x_{2},q_{\perp}\right)=C(x_{1},x_{2})\,\sum_{j=0}^{2}c_{j}\,\left\{ -\ln\frac{\kappa_{j}}{\kappa}-\left[4\frac{m^{2}}{q_{\perp}^{2}}-2\frac{m_{\pi}^{2}}{q_{\perp}^{2}}x_{1}(1-x_{1})+1\right]f\left(\frac{\kappa_{j}}{q_{\perp}^{2}}\right)\right\} \,,\label{aa}
\end{equation}
\begin{equation}
F_{u,\Delta\bar{d}}^{\left(PV\right)}\left(x_{1},x_{2},q_{\perp}\right)=F_{\Delta u,\bar{d}}^{\left(PV\right)}\left(x_{1},x_{2},q_{\perp}\right)=0\label{ua}
\end{equation}
\begin{equation}
F_{\delta u,\delta\bar{d}}^{s\left(PV\right)}\left(x_{1},x_{2},q_{\perp}\right)=-C(x_{1},x_{2})\,\sum_{j=0}^{2}c_{j}\,2\frac{m^{2}}{q_{\perp}^{2}}f\left(\frac{\kappa_{j}}{q_{\perp}^{2}}\right)\,,\label{tt0}
\end{equation}
\begin{equation}
F_{\delta u,\delta\bar{d}}^{t\left(PV\right)}\left(x_{1},x_{2},q_{\perp}\right)=C(x_{1},x_{2})\,\sum_{j=0}^{2}c_{j}\,2\frac{\kappa_{j}}{q_{\perp}^{2}}f\left(\frac{\kappa_{j}}{q_{\perp}^{2}}\right)\,,\label{tt2}
\end{equation}
\begin{equation}
F_{u,\delta\bar{d}}^{v\left(PV\right)}\left(x_{1},x_{2},q_{\perp}\right)=-F_{\delta u,\bar{d}}^{v\left(PV\right)}\left(x_{1},x_{2},q_{\perp}\right)=-C(x_{1},x_{2})\,\sum_{j=0}^{2}c_{j}\,2\frac{m}{q_{\perp}}f\left(\frac{\kappa_{j}}{q_{\perp}^{2}}\right)\,.\label{vt}
\end{equation}
\begin{equation}
F_{\Delta u,\delta\bar{d}}^{v\left(PV\right)}\left(x_{1},x_{2},q_{\perp}\right)=-F_{\delta u,\Delta\bar{d}}^{v\left(PV\right)}\left(x_{1},x_{2},q_{\perp}\right)=0\label{at}
\end{equation}
where $\kappa= m^2-m_\pi^2x_1(1-x_1)$, $\kappa_j= M_j^2-m_\pi^2x_1(1-x_1)$
and
\begin{eqnarray}
C(x_1,x_2) & = &
\left(\frac{N_{c}\,g_{\pi qq}^{2}}{4\pi^{2}}\right)
\delta\left(x_{1}+x_{2}-1\right)\,\theta\left(x_{1}\right)\,\theta
\left(1-x_{1}\right) \, ,\label{A.C}
\\
f(a) & = & \frac{1}{\sqrt{4a + 1}}\, \log \frac
{\sqrt{4a +1}+1}{\sqrt{4a +1}-1} \quad.\label{f_a}
\end{eqnarray}

 In the case of Eqs.~(\ref{ua}) and~(\ref{at}), the traces involved
in Eq.~(\ref{A.11}) are linear in the integrated momentum, $\vec{k}_{\perp},$
and, therefore, the corresponding dPDFs vanish after the integration present
in Eq.~(\ref{A.10}). In ref.~\cite{Diehl1}, as reported before in this section, the result of Eqs.~(\ref{ua}) is predicted according to parity arguments.

One should notice first of all that, as expected by the  Gaunt sum rule at $q_{\perp}^2=0$ \cite{gaunt}, the $x_2$ integral yields
the expression for the parton distribution function (PDF) reported,
in the same NJL framework with Pauli-Villars 
%\textcolor{red}{(¿and with the Light Front?)} 
regularization,
in Refs.~\cite{Theussl:2002xp,Courtoythesis}.

We are now in the position to show our results for the pion dPDF for  the $\pi^+$, in the NJL model.
The dashed lines
in Fig.~\ref{Fig_dPDF_q} represent the quantities
\begin{eqnarray}
F_{a_1\bar{a}_2}^{(PV)}\left( {q}_{\perp} \right) =
\int d x_1 d x_2
F_{a_1\bar{a}_2}^{(PV)}\left(x_{1},x_{2}, {q}_{\perp}
\right) \, ,
\label{inqt}
\end{eqnarray}
where
$F_{a_1\bar{a_2}}^{\left({PV}\right)}\left(x_{1},x_{2}, {q}_{\perp}
\right)$ 
is the generic scalar function used to define
the momentum space dPDFs in Eqs. (\ref{TensorStr_q}).

\begin{figure}
\begin{center}
\includegraphics[scale=0.5]{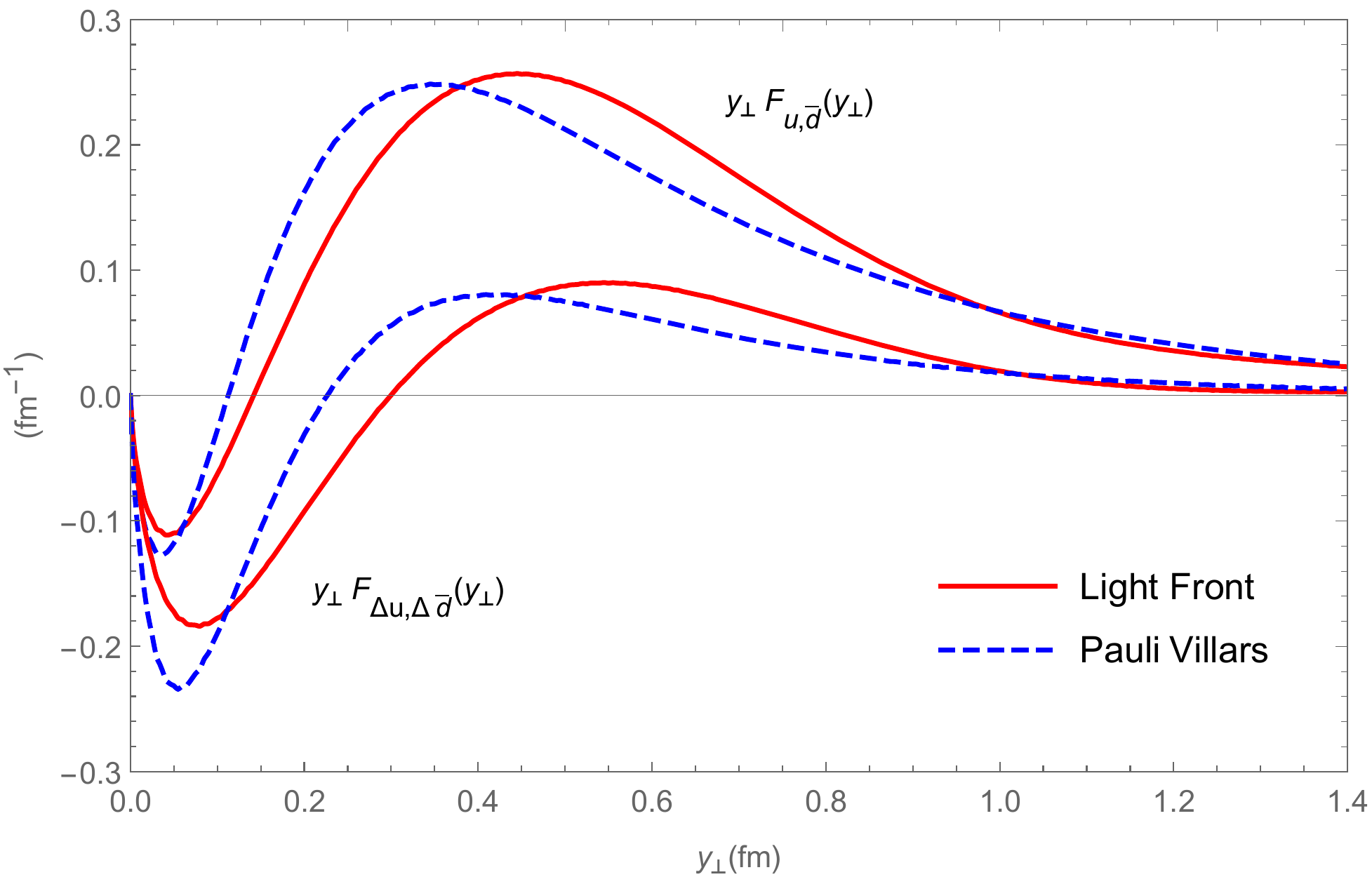} 
\vskip 4.mm
\includegraphics[scale=0.5]{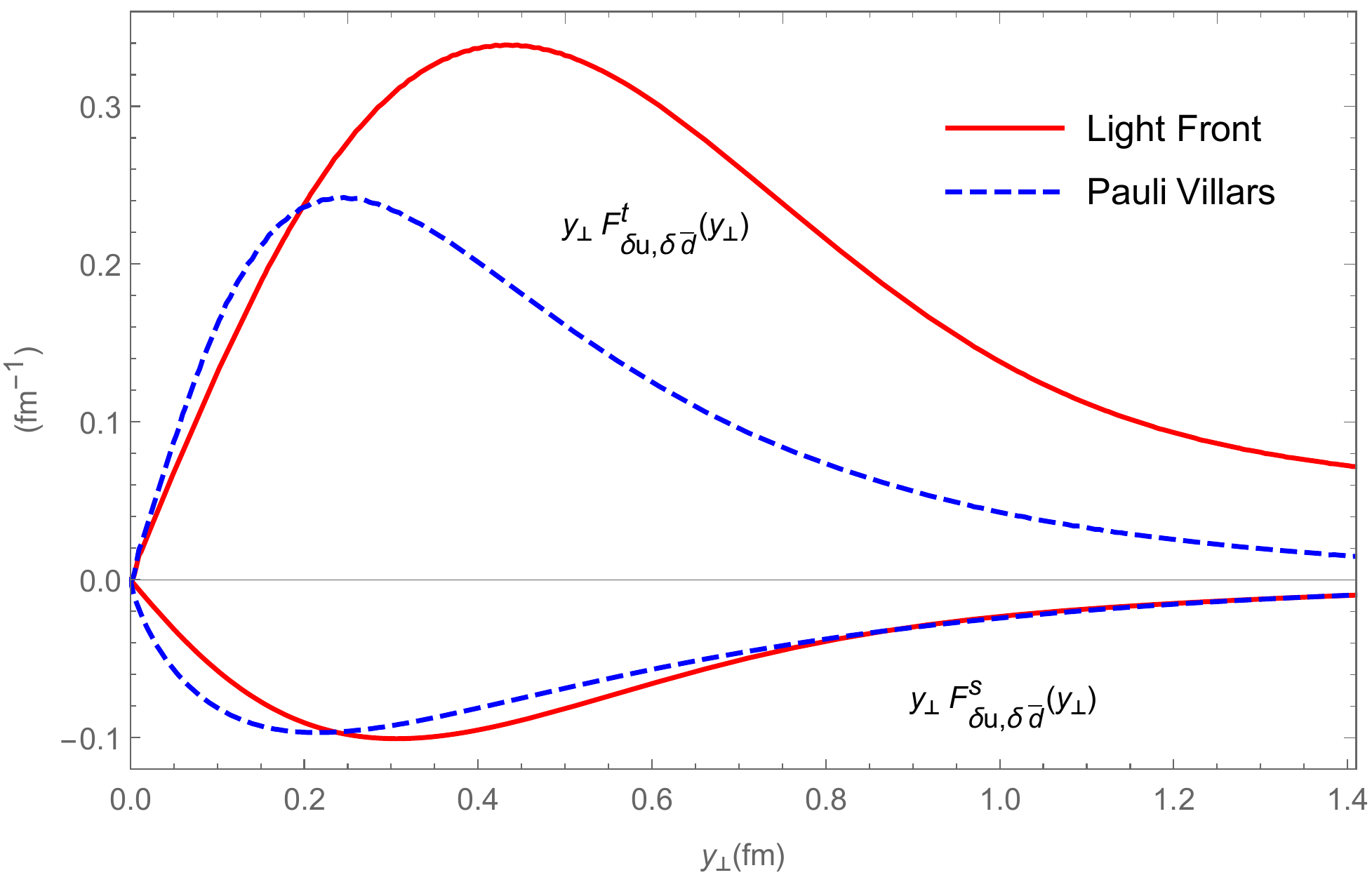}
\vskip 4.mm
\includegraphics[scale=0.5]{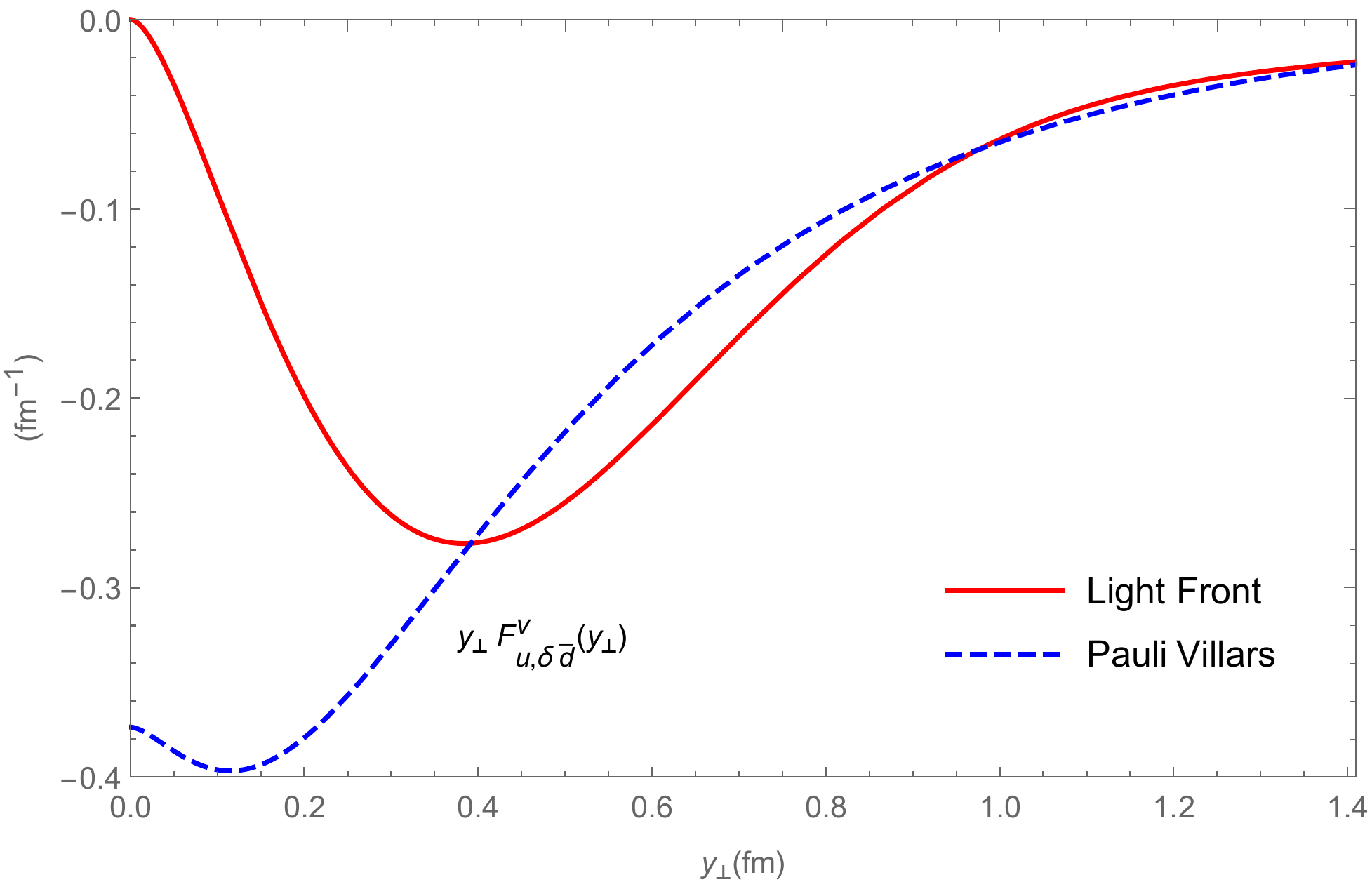}
\caption{As in Fig. \ref{Fig_dPDF_q}, but in transverse coordinate space}
\label{Fig_dPDF_y}

\end{center}

\end{figure}

We show
the results also in coordinate space, obtained 
by integrating
over $x_1,x_2$ the scalar functions
defined in
Eqs. (\ref{fts})-
(\ref{ftt}).
%Fourier transforming  Eq.~\eqref{inqt}:
%\begin{eqnarray}
%F_{u\bar{d}}^{\left(0\right)}\left( {y}_{\perp} \right) =
%\frac{1}{(2 \pi)^2}
%\int d^2 q_\perp e^{-i \vec y_\perp \cdot \vec q_\perp}
%F_{u\bar{d}}^{\left(0\right)}\left( {q}_{\perp}
%\right) \, .
%\label{inyt}
%\end{eqnarray}
These quantities, multiplied by $y_\perp$, 
are given by the dashed
lines in Fig.~\ref{Fig_dPDF_y}.
%within our
%regularization scheme and also using that proposed
%in Ref.~\cite{Itakura:2000te},  previously introduced, 
%with the same convention for the curves and their colors.
%The result obtained within the regularization scheme 
%of Ref.~\cite{Davidson:2001cc} cannot be distinguished
% from our result and it is not shown.
%Again, the difference related to the alternative choices
%of regularization is rather small.

A few relevant comments are in order. 

As stated in the Introduction, the scalar quantity
$F_{u,\bar{d}}^{\left(PV\right)}\left(y_{\perp}\right)$,
should represent, in principle, the probability density
to have the two particles at a given transverse distance
$y_\perp$. Indeed, our result for this function is properly normalized,
as it can be read from Fig. \ref{Fig_dPDF_q}
($F_{u \bar d}^{\left(PV\right)}
\,(q_\perp=0)=1$).
Nonetheless it is found that $F_{u \bar d}^{\left(PV\right)}(y_\perp)$
turns out to be negative at low values of $y_\perp$.

Another peculiar-looking feature of the results is related to the function 
$F_{u,\delta \bar{d}}^{v}\left(q_{\perp}
\right)$. This quantity presents a { slowly decreasing} tail at high values of $q_\perp$, which
produces in coordinate space a peculiar behavior at low values of $y_\perp$.

%These two facts could appear as drawbacks of the model,which has to be regularized using a subtraction procedure.
{As these features  are model dependent,
one could wonder 
to what extent they are affected
by the choice of the regularization scheme, as part of the model.} 
For this reason we have performed the calculation using another,
novel regularization procedure, suitable for calculation involving light-cone variables.
This method, called here after
Light Front (LF) regularization, is carefully
described in Appendix \ref{App.NJL_LF_regularization}. Some
intermediate results are included in 
Appendix \ref{App.Intermediate-Results}.

Our final results in the LF scheme are
\begin{equation}
F_{u,\bar{d}}^{\left(LF\right)}\left(x_{1},x_{2},q_{\perp}\right)=\tilde{C}(x_{1},x_{2})\,\left[g_{2}\left(q_{\perp}\right)+\left(\frac{\tilde{m}^{2}}{q_{\perp}^{2}}-\frac{1}{4}\right)\,g_{0}\left(q_{\perp}\right)\right]\,,\label{A.16-1}
\end{equation}
\begin{equation}
F_{\Delta u,\Delta\bar{d}}^{\left(LF\right)}\left(x_{1},x_{2},q_{\perp}\right)=\tilde{C}(x_{1},x_{2})\,\left[g_{2}\left(q_{\perp}\right)-\left(\frac{\tilde{m}^{2}}{q_{\perp}^{2}}+\frac{1}{4}\right)\,g_{0}\left(q_{\perp}\right)\right]\,,\label{aa-1}
\end{equation}
\begin{equation}
F_{u,\Delta\bar{d}}^{\left(LF\right)}\left(x_{1},x_{2},q_{\perp}\right)=F_{\Delta u,\bar{d}}^{\left(LF\right)}\left(x_{1},x_{2},q_{\perp}\right)=0\label{ua-1}
\end{equation}
\begin{equation}
F_{\delta u,\delta\bar{d}}^{s\left(LF\right)}\left(x_{1},x_{2},q_{\perp}\right)=-\tilde{C}(x_{1},x_{2})\,\frac{\tilde{m}^{2}}{q_{\perp}^{2}}\,g_{0}\left(q_{\perp}\right)\,,\label{tt0-1}
\end{equation}
\begin{equation}
F_{\delta u,\delta\bar{d}}^{t\left(LF\right)}\left(x_{1},x_{2},q_{\perp}\right)=-\tilde{C}(x_{1},x_{2})\,\left[g_{2}\left(q_{\perp}\right)-2\,\tilde{g}_{2}\left(q_{\perp}\right)+\frac{1}{4}\,g_{0}\left(q_{\perp}\right)\right]\,,\label{tt2-1}
\end{equation}
\begin{equation}
F_{u,\delta\bar{d}}^{v\left(LF\right)}\left(x_{1},x_{2},q_{\perp}\right)=-F_{\delta u,\bar{d}}^{v\left(LF\right)}\left(x_{1},x_{2},q_{\perp}\right)=-\tilde{C}(x_{1},x_{2})\,\frac{\tilde{m}}{q_{\perp}}\,g_{0}\left(q_{\perp}\right)\,.\label{vt-1}
\end{equation}
\begin{equation}
F_{\Delta u,\delta\bar{d}}^{v\left(LF\right)}\left(x_{1},x_{2},q_{\perp}\right)=-F_{\delta u,\Delta\bar{d}}^{v\left(LF\right)}\left(x_{1},x_{2},q_{\perp}\right)=0\label{at-1}
\end{equation}
with $\tilde{\kappa}=\tilde{m}^{2}-m_{\pi}^{2}x_{1}(1-x_{1})$ and
\begin{equation}
\tilde{C}(x_1,x_2) = 
\left(\frac{N_{c}\,\tilde{g}_{\pi qq}^{2}}{4\pi^{2}}\right)
\delta\left(x_{1}+x_{2}-1\right)\,\theta\left(x_{1}\right)\,\theta
\left(1-x_{1}\right) \, ,\label{A.C.1}
\end{equation}
\begin{equation}
g_{0}\left(q_{\perp}\right)=2\sqrt{\frac{q_{\perp}^{2}}{4\tilde{\kappa}+q_{\perp}^{2}}}\log\left(\frac{\left(\sqrt{4\tilde{\kappa}+q_{\perp}^{2}}+\sqrt{q_{\perp}^{2}}\right)\left(4\tilde{\kappa}+4\Lambda_{\perp}^{2}+q_{\perp}^{2}\right)}{\sqrt{q_{\perp}^{2}}\left(4\tilde{\kappa}-4\Lambda_{\perp}^{2}+q_{\perp}^{2}\right)+\sqrt{4\tilde{\kappa}+q_{\perp}^{2}}\sqrt{\left(4\tilde{\kappa}+4\Lambda_{\perp}^{2}-q_{\perp}^{2}\right)^{2}+16\tilde{\kappa}q_{\perp}^{2}}}\right)\,,
\end{equation}
\begin{equation}
g_{2}\left(q_{\perp}\right)=-\frac{4\tilde{\kappa}+q_{\perp}^{2}}{4q_{\perp}^{2}}g_{0}\left(q_{\perp}\right)+\log\left(\frac{4\tilde{\kappa}+4\Lambda_{\perp}^{2}+\sqrt{\left(4\tilde{\kappa}+4\Lambda_{\perp}^{2}-q_{\perp}^{2}\right)^{2}+16\tilde{\kappa}q_{\perp}^{2}}-q_{\perp}^{2}}{8\tilde{\kappa}}\right)\,,
\end{equation}
\begin{align}
\tilde{g}_{2}\left(q_{\perp}\right) & =\frac{-4\tilde{\kappa}-4\Lambda_{\perp}^{2}+\sqrt{\left(4\tilde{\kappa}+4\Lambda_{\perp}^{2}-q_{\perp}^{2}\right)^{2}+16\tilde{\kappa}q_{\perp}^{2}}-q_{\perp}^{2}}{4q_{\perp}^{2}}\nonumber \\
 & +\frac{1}{2}\log\left(\frac{4\tilde{\kappa}+4\Lambda_{\perp}^{2}+\sqrt{\left(4\tilde{\kappa}+4\Lambda_{\perp}^{2}-q_{\perp}^{2}\right)^{2}+16\tilde{\kappa}q_{\perp}^{2}}-q_{\perp}^{2}}{8\tilde{\kappa}}\right)\,.
\label{A.C.3}
\end{align}

As it happens in the $PV$ regularization scheme,
in the case of Eqs.~(\ref{ua-1}) and~(\ref{at-1}) the traces involved
in Eq.~(\ref{A.11}) are linear in the integrated momentum, $\vec{k}_{\perp},$
and, therefore, the corresponding dPDF vanish after the integration present
in Eq. (\ref{A.10}). In ref. \cite{Diehl1}, as reported before in this section, the result of Eq. (\ref{ua-1}) is predicted according to parity conservation.

As in the $PV$ case
one should notice first of all that, as expected by the  Gaunt sum rule at $q_{\perp}^2=0$ \cite{gaunt}, the $x_2$ integral provides 
the expression for the parton distribution function (PDF) obtained using the same
regularization.

In Figs. \ref{Fig_dPDF_q} and
\ref{Fig_dPDF_y} the results obtained
in the LF regularization scheme,
given by full lines,
are compared to those presented in
the $PV$ case.

It is found that, for
$q_{\perp}<0.5$ GeV, all the distributions
have a very similar behavior using the
$PV$ or the LF regularization.
At higher values of $q_{\perp}$ this
is still true for the functions
$F_{u,\bar{d}},\,F_{\Delta u,\Delta\bar{d}},$ and $F_{\delta u,\delta\bar{d}}^{s}$,
while a sizable difference is found for $F_{\delta u,\delta\bar{d}}^{t}$ and $F_{u,\delta\bar{d}}^{v}$.
In general, with increasing
$q_{\perp}$, the distributions 
regularized within the LF method decrease
faster than those regularized using $PV$.
This is evident in particular for
$F_{u,\delta\bar{d}}^{v}.$
To have a quantitative flavor of this trend,
the behavior of all the distributions, when
$q_\perp \rightarrow 0$
and
when $q_\perp \rightarrow \infty$,
are summarized in Tables 2 and 3, { in the chiral limit,
for the two methods of regularization used.
Results with physical masses would differ by a few percent}.

In coordinate space we have that
the choice of regularization does
not affect strongly the distributions
$F_{u,\bar{d}},\,F_{\Delta u,\Delta\bar{d}},$ and $F_{\delta u,\delta\bar{d}}^{s}$.
This is not the case for
$F_{\delta u,\delta\bar{d}}^{t}$ and
for $F_{u,\delta\bar{d}}^{v}$.
Predictions differ in the first case
at high values of $y_{\perp}$,
while in the second case
they differ at low $y_{\perp}$.
In the latter situation, the behavior
of the result regularized in the LF
scheme, related to a fast
decrease of the corresponding distribution
in momentum space, appears more natural
than the peculiar one found using $PV$,
previously described.

Let us see if the other peculiar-looking
trend observed using $PV$ regularization,
i.e. the presence of a tiny region of negative $F_{u \bar d}(y_\perp)$,
is found also in LF regularization.
Actually, we find that this feature
is rather independent
on the regularization and arises in both
schemes.
The origin of this region of negative 
$F_{u \bar d}(y_\perp)$ could be actually more general.
A warning on the issue of positivity
of dPDFs, even in the unpolarized, vector case of interest
here, has been reported in Ref.~\cite{Diehl:2013mla}.
The interpretation of any parton distribution as a probability
density is not strictly valid 
in QCD, because the distributions are defined with 
subtractions from the ultraviolet
region of parton momenta, which can invalidate
their positivity. This is mostly true
beyond leading order:
it is well known that, starting at NLO, a dependence on
the factorization scheme is found and the probability interpretation is lost, even for PDFs. In the present field theoretical approach, where a given order
cannot be assigned unambiguously to the calculation,
the interpretation of the dPDF as a probability density
is questionable and the result we obtain is
less surprising than what it could seem at a first sight.

%In any case, this peculiar behavior at low values of $y$ is not very relevant for phenomenological considerations.
%Our calculation is performed in momentum space and the coordinate space result is obtained through Fourier transformation. 
The negative yield
at very low $y$ values is a consequence of the long negative tail
found in $q_\perp$ space (see Fig. \ref{Fig_dPDF_q}) {and  is} actually not relevant phenomenologically, since
the dPDF and the associated DPS cross section at high $q_\perp$ is expected to be very small.
As a matter of facts,
in the rest of the paper we will not show results at $q_\perp$ larger
than 0.5 GeV.

{ We note in passing that $q_\perp$ is a momentum unbalance,
not related to the internal pion dynamics but rather to
the insertion of an external momentum.
Interestingly, we have found that the introduction of a properly chosen $q_{\perp}$-dependent cut-off in evaluating Eqs.~\eqref{A.C.1}-
\eqref{A.C.3} removes all the negative values of $F_{u \bar d}(y_\perp)$, suggesting that a momentum dependent procedure
might be motivated in the present
situation.}

\begin{table}[t]
\begin{centering}
\begin{tabular}{|c|c|c|}
\hline 
 & PV & LF\tabularnewline
\hline 
$\int dx_{1}dx_{2}F_{u,\bar{d}}\left(x_{1},x_{2},q_{\perp}\right)$ & $1-3.98\,\text{\ensuremath{q_{\perp}^{2}}}$ & $1-4.05\,\text{\ensuremath{q_{\perp}^{2}}}$\tabularnewline
\hline 
$\int dx_{1}dx_{2}F_{\Delta u,\Delta\bar{d}}\left(x_{1},x_{2},q_{\perp}\right)$ & $0.10-1.04\,\text{\ensuremath{q_{\perp}^{2}}}$ & $0.09-1.09\,\text{\ensuremath{q_{\perp}^{2}}}$\tabularnewline
\hline 
$\int dx_{1}dx_{2}F_{\delta u,\delta\bar{d}}^{s}\left(x_{1},x_{2},q_{\perp}\right)$ & $-0.45+1.47\,\text{\ensuremath{q_{\perp}^{2}}}$ & $-0.45+1.48\,\text{\ensuremath{q_{\perp}^{2}}}$\tabularnewline
\hline 
$\int dx_{1}dx_{2}F_{\delta u,\delta\bar{d}}^{t}\left(x_{1},x_{2},q_{\perp}\right)$ & $-1.33\,\text{\ensuremath{q_{\perp}^{2}}}+5.18\,\text{\ensuremath{q_{\perp}^{4}}}$ & $-1.43\,\text{\ensuremath{q_{\perp}^{2}}}+4.91\,\text{\ensuremath{q_{\perp}^{4}}}$\tabularnewline
\hline 
$\int dx_{1}dx_{2}F_{u,\delta\bar{d}}^{v}\left(x_{1},x_{2},q_{\perp}\right)$ & $-\sqrt{\text{\ensuremath{q_{\perp}^{2}}}}\,\left[1.89-6.17\,\text{\ensuremath{q_{\perp}^{2}}}\right]$ & $-\sqrt{\text{\ensuremath{q_{\perp}^{2}}}}\,\left[1.85-6.01\,\text{\ensuremath{q_{\perp}^{2}}}\right]$\tabularnewline
\hline 
\end{tabular}\caption{Behavior of
the function
$\int dx_{1}dx_{2}F_{a_{1},\bar{a}}\left(x_{1},x_{2},q_{\perp}\right)$, for
$q_{\perp}\rightarrow0$, in the
chiral limit.}
\par\end{centering}
\end{table}

\begin{table}[t]
\centering{}%
\begin{tabular}{|c|c|c|}
\hline 
 & PV & LF\tabularnewline
\hline 
$\int dx_{1}dx_{2}F_{u,\bar{d}}\left(x_{1},x_{2},q_{\perp}\right)$ & $-0.87\,\text{\ensuremath{q_{\perp}^{-2}}}$ & $-0.71\,\text{\ensuremath{q_{\perp}^{-2}}}$\tabularnewline
\hline 
$\int dx_{1}dx_{2}F_{\Delta u,\Delta\bar{d}}\left(x_{1},x_{2},q_{\perp}\right)$ & $-1.09\,\text{\ensuremath{q_{\perp}^{-2}}}$ & $-0.71\,\text{\ensuremath{q_{\perp}^{-2}}}$\tabularnewline
\hline 
$\int dx_{1}dx_{2}F_{\delta u,\delta\bar{d}}^{s}\left(x_{1},x_{2},q_{\perp}\right)$ & $-0.11\,\text{\ensuremath{q_{\perp}^{-2}}}$ & $-0.17\,\text{\ensuremath{q_{\perp}^{-4}}}$\tabularnewline
\hline 
$\int dx_{1}dx_{2}F_{\delta u,\delta\bar{d}}^{t}\left(x_{1},x_{2},q_{\perp}\right)$ & $-0.87\,\text{\ensuremath{q_{\perp}^{-2}}}$ & $-0.71\,\text{\ensuremath{q_{\perp}^{-2}}}$\tabularnewline
\hline 
$\int dx_{1}dx_{2}F_{u,\delta\bar{d}}^{v}\left(x_{1},x_{2},q_{\perp}\right)$ & $-0.48\,\text{\ensuremath{q_{\perp}^{-1}}}$ & $-0.69\,\text{\ensuremath{q_{\perp}^{-3}}}$\tabularnewline
\hline 
\end{tabular}\caption{
Behavior of
the function
$\int dx_{1}dx_{2}F_{a_{1},\bar{a}}\left(x_{1},x_{2},q_{\perp}\right)$, for
$q_{\perp}\rightarrow \infty$, in the
chiral limit.
}
\end{table}

%As a final consideration we remark that, in the present NJL scheme,
%a two-hard scattering mechanism such as the one depicted  in Fig.
%\ref{Fig_two_GPDs} for a $\pi^{+}$,
%with two bilocal currents acting, one associated to the quark $u$ and the other to the antiquark $\bar{d}$,
%has to be considered
%and could give a contribution to pion dPDFs.
%We have therefore analyzed this diagram, and the other
%in which the first current acts on the antiquark and 
%the second on the quark. After a careful analysis, it turns out
%that, due to the structure of the propagators
%of the interacting partons, any possible contribution of this kind 
%to pion dDPDFs is actually vanishing.

\begin{figure}
\begin{centering}
\includegraphics[scale=0.6]{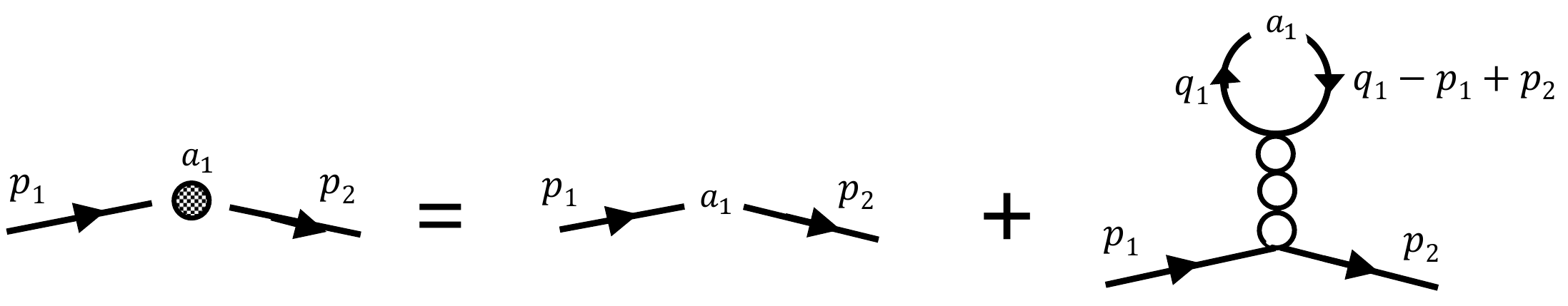}
\par\end{centering}
\caption{Dressed vertex associated to the bilocal currents}
\label{Fig_Dressed_vertex} 
\end{figure}

\subsection{Dressing the bilocal vertex and other contributions in the NJL model}

In addition to the contribution Eq.~\eqref{A.7},
in the NJL model we must consider also
the dressing of the bilocal vertex
due to the chiral interaction. This corresponds to change the bare vertex
given in Table~\ref{Tab_bare_vertex} by the dressed one depicted
in Fig.~\ref{Fig_Dressed_vertex}. Therefore, instead of using $\bar{\Gamma}_{a_{1}}\,\delta\left(P^{+}x_{1}-\frac{1}{2}\left(p_{1}+p_{2}\right)^{+}\right)$
for the bare vertex in the Feynman amplitude in Eq.~(\ref{A.9}), we must use the replacement
%
%\begin{align}
%\bar{\Gamma}_{a_{1}}\,\delta\left(P^{+}x_{1}-\frac{1}{2}\left(p_{1}+p_{2}\right)^{+}\right)\longrightarrow & \bar{\Gamma}_{a_{1}}\,\delta\left(P^{+}x_{1}-\frac{1}{2}\left(p_{1}+p_{2}\right)^{+}\right)+\nonumber \\
%\frac{2ig}{1-2g\,\Pi_{S}\left(q^{2}\right)}\left(-\right)\int\frac{d^{4}q_{1}}{\left(2\pi\right)^{4}} & \delta\left(x_{1}P^{+}-q_{1}^{+}\right)\text{Tr}\left[iS_{F}\left(q_{1}\right)\,iS_{F}\left(q_{1}-q\right)\,\bar{\Gamma}_{a_{1}}\right]+\nonumber \\
%\sum_{j=1}^{3}i\gamma_{5}\tau^{j}\frac{2ig}{1-2g\,\Pi_{PS}\left(q^{2}\right)}\left(-\right)\int\frac{d^{4}q_{1}}{\left(2\pi\right)^{4}} & \delta\left(x_{1}P^{+}-q_{1}^{+}\right)\text{Tr}\left[iS_{F}\left(q_{1}\right)\,i\gamma_{5}\tau^{j}\,iS_{F}\left(q_{1}-q\right)\,\bar{\Gamma}_{a_{1}}\right]\label{A.17} \, ,
%\end{align}
%

%
\begin{eqnarray}
&&\bar{\Gamma}_{a_{1}}\,\delta\left(P^{+}x_{1}-\frac{1}{2}\left(p_{1}+p_{2}\right)^{+}\right)\nonumber\\
\longrightarrow && \bar{\Gamma}_{a_{1}}\,\delta\left(P^{+}x_{1}-\frac{1}{2}\left(p_{1}+p_{2}\right)^{+}\right)\nonumber \\
&&+\frac{2ig}{1-2g\,\Pi_{S}\left(q^{2}\right)}\left(-\right)\int\frac{d^{4}q_{1}}{\left(2\pi\right)^{4}}  \delta\left(x_{1}P^{+}-q_{1}^{+}\right)\text{Tr}\left[iS_{F}\left(q_{1}\right)\,iS_{F}\left(q_{1}-q\right)\,\bar{\Gamma}_{a_{1}}\right]\nonumber \\
&&+\sum_{j=1}^{3}i\gamma_{5}\tau^{j}\frac{2ig}{1-2g\,\Pi_{PS}\left(q^{2}\right)}\left(-\right)\int\frac{d^{4}q_{1}}{\left(2\pi\right)^{4}}  \delta\left(x_{1}P^{+}-q_{1}^{+}\right)\text{Tr}\left[iS_{F}\left(q_{1}\right)\,i\gamma_{5}\tau^{j}\,iS_{F}\left(q_{1}-q\right)\,\bar{\Gamma}_{a_{1}}\right] \, ,\nonumber\\
\label{A.17}
\end{eqnarray}
where $q=p_{1}-p_{2}.$ Here, $\Pi_{S}$ and $\Pi_{PS}$ are the scalar
and pseudo-scalar polarizations, respectively,
defined in Appendix \ref{App.NJL_Basic_Equations}.

Performing the explicit calculation of the dressing term, we found
that it vanishes. Effectively, due to the fact that $q^{+}=0,$ the
integrals over $q_{1}$ present in Eq. (\ref{A.17}) are\\% proportionals
%to \\
%(dos alternativas para la ecuación. Opcion A)
%\[
%\int\frac{dq_{1}^{-}}{2\pi}\,\frac{%f\left(q_{1}q\right)}{\left(2\,x_{1}P^{+}\right)^{2}\left(q_{1}^{-}-\frac{\vec{q}_{1\perp}^{2}+m^{2}}{2\,x_{1}P^{+}}+i\frac{\epsilon}{2\,x_{1}P^{+}}\right)\left(q_{1}^{-}-q^{-}-\frac{\left(\vec{q}_{1\perp}-\vec{q}_{\perp}\right)^{2}+m^{2}}{2\,x_{1}P^{+}}+i\frac{\epsilon}{2\,x_{1}P^{+}}\right)}=0\,\,,
%\]
%(Opcion B)
\begin{align}
\int\frac{d^{4}q_{1}}{\left(2\pi\right)^{4}}\delta\left(x_{1}P^{+}-q_{1}^{+}\right)\,\text{Tr}\left[iS_{F}\left(q_{1}\right)\left(1,i\gamma_{5}\right)iS_{F}\left(q_{1}-q\right)\bar{\Gamma}_{a_{1}}\right] & \propto\nonumber \\
\int\frac{dq_{1}^{-}}{2\pi}\,\frac{f\left(q_{1},q\right)}{\left(2\,x_{1}P^{+}\right)^{2}\left(q_{1}^{-}-\frac{\vec{q}_{1\perp}^{2}+m^{2}}{2\,x_{1}P^{+}}+i\frac{\epsilon}{2\,x_{1}P^{+}}\right)\left(q_{1}^{-}-q^{-}-\frac{\left(\vec{q}_{1\perp}-\vec{q}_{\perp}\right)^{2}+m^{2}}{2\,x_{1}P^{+}}+i\frac{\epsilon}{2\,x_{1}P^{+}}\right)} & =0\,\,,\label{No se sabe}
\end{align}
being $f\left(q_{1},q\right)$ some function of $q^{-},$ $q_{\perp}$,
$q_{1}^{-}$and $\vec{q}_{1\perp}.$ The last integral vanishes due
to the fact that the poles of both propagators are in the same half
complex plane. Therefore, the dressing of the bilocal vertex does
not give any additional contribution to all the different pion dPDFs.

\begin{figure}
%\begin{centering}
\includegraphics[scale=0.5]{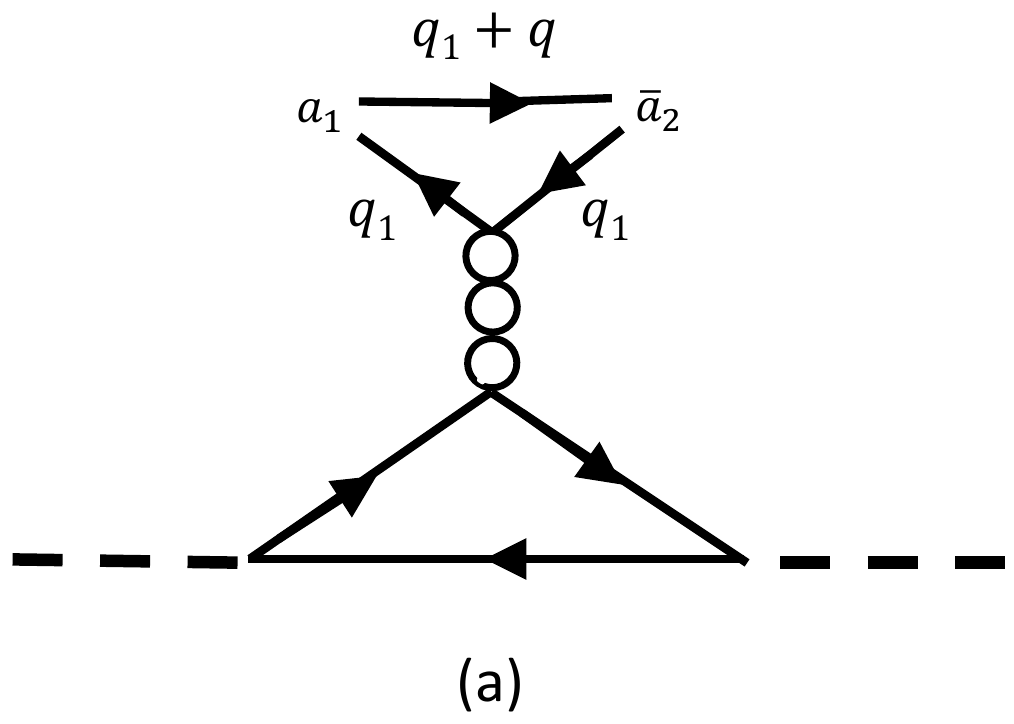}~~~~~~~~\includegraphics[scale=0.5]{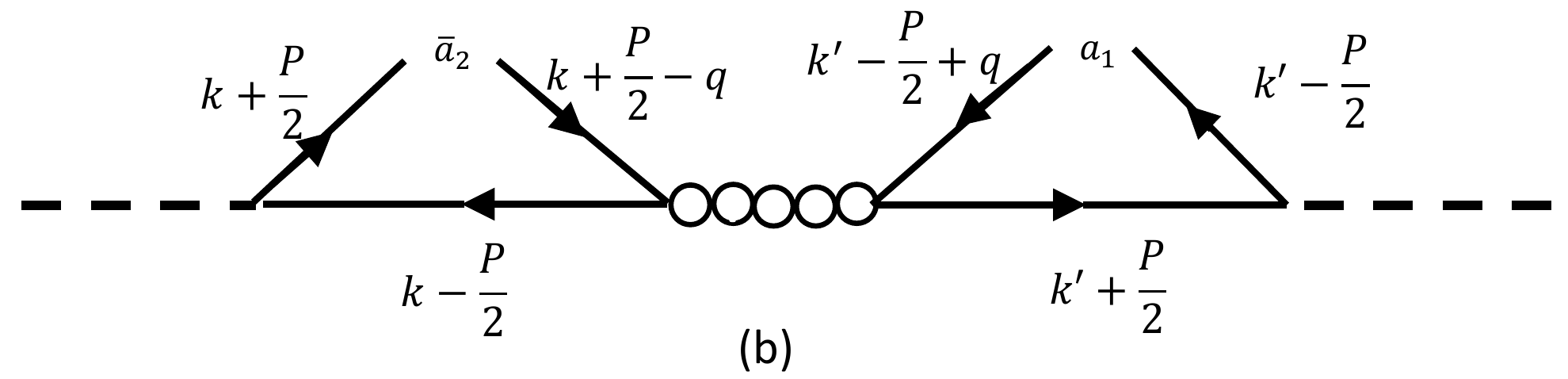}
\caption{Other possible diagrams in the NJL model. All of them do not contribute { to the dPDFs}, as explained
in the text.}
\label{Other_Contrib_Diagrams}
%\end{centering}
\end{figure}

Two other types of possible contributions to the dPDFs, depicted in
Fig \ref{Other_Contrib_Diagrams}, have to be considered. Despite the
fact that, apparently, these two contributions are higher order contributions,
chiral symmetry, present in the NJL model, 
{ensures}
that they are of the same order than the one depicted in Fig.~\ref{Fig_diagrama_rombo}.

The type of contribution shown in Fig.~\ref{Other_Contrib_Diagrams}a represents
the possibility that the two involved partons are originated by the
same vacuum fluctuation, which, in order to make a connected diagram, has to be
connected to the pion line. To analyze this kind of diagrams, we
focus our attention on the upper triangle. In the particular case
of the diagram of Fig.~\ref{Other_Contrib_Diagrams}a we have, for this
part,
\begin{align}
\int\frac{d^{4}q_{1}}{\left(2\pi\right)^{4}}\delta\left(x_{2}P^{+}+q_{1}^{+}\right)\delta\left(x_{1}P^{+}-q_{1}^{+}\right)\,\text{Tr}\left[iS_{F}\left(q_{1}\right)\bar{\Gamma}_{a_{1}}iS_{F}\left(q_{1}+q\right)\bar{\Gamma}_{\bar{a}_{2}}iS_{F}\left(q_{1}\right)\right] & \propto\nonumber \\
\int\frac{dq_{1}^{-}}{2\pi}\,\frac{g\left(q_{1},q\right)}{\left(2\,x_{1}P^{+}\right)^{3}\left(q_{1}^{-}-\frac{\vec{q}_{1\perp}^{2}+m^{2}}{2\,x_{1}P^{+}}+i\frac{\epsilon}{2\,x_{1}P^{+}}\right)^{2}\left(q_{1}^{-}+q^{-}-\frac{\left(\vec{q}_{1\perp}+\vec{q}_{\perp}\right)^{2}+m^{2}}{2\,x_{1}P^{+}}+i\frac{\epsilon}{2\,x_{1}P^{+}}\right)} & =0\,\,\label{No se sabe-1} \, ,
\end{align}
being $g\left(q_{1},q\right)$ some function of $q^{-},$ $q_{\perp}$,
$q_{1}^{-}$ and $\vec{q}_{1\perp}.$ As in the case of Eq.~(\ref{No se sabe}),
the last integral vanishes because all the poles of the propagators
are in the same half complex plane. Therefore, diagrams of type of
Fig.~\ref{Other_Contrib_Diagrams}a give no contribution to the dPDFs.
With respect to this, it is more interesting
to see what happens with contributions of the type shown in Fig.~\ref{Other_Contrib_Diagrams}b.
Despite of their aspect, this kind of contributions are not related
to {the approximation to dPDF in terms of } GPDs {proposed
in Refs.~\cite{blok1} and~\cite{blok2}}. First of all, the intermediate state represented by
the small bubbles has the quantum numbers of a pion. This part of the
diagram gives a contribution which, close
to the pion mass,
can be approximated by the point-like pion propagator, 
\begin{equation}
\frac{2ig_{0}}{1-2g_{0}\Pi_{PS}\left(\left(P-q\right)^{2}\right)}\simeq\frac{-i\,g_{\pi qq}^{2}}{\left(P-q\right)^{2}-m_{\pi}^{2}} \, .
\end{equation}
According to Eq.~(\ref{A.9}), all possible diagrams have an overall
$q^{-}$-integral. In the case of the present diagram, this integral
only involves the pion propagator and two quark propagators, one from
each of the triangles. Extracting these terms we obtain (here $h\left(k,\,k^{\prime},\,\vec{q}_{\perp},\,q^{-}\right)$
is some function of $q^{-},$ $q_{\perp}$, $k^{-},$ $\vec{k}_{\perp},$
$k^{\prime-},$ and $\vec{k}_{\perp}^{\prime}$)
\begin{eqnarray}
&&\int\frac{dq^{-}}{2\pi}\frac{h\left(k,\,k^{\prime},\,\vec{q}_{\perp},\,q^{-}\right)}{\left[\left(P-q\right)^{2}-m_{\pi}^{2}+i\epsilon\right]\left[\left(k+\frac{P}{2}-q\right)^{2}-m^{2}+i\epsilon\right]\left[\left(k^{\prime}-\frac{P}{2}+q\right)^{2}-m_{\pi}^{2}+i\epsilon\right]}\nonumber \\
&&=\frac{-1}{\left(2P^{+}\right)^{3}\,x_{1}\,x_{2}}\label{complicated} \\
&&\left\{ \frac{i\,\theta\left(-x_{1}\right)\,\theta\left(-x_{2}\right)\,h\left(k,\,k^{\prime},\,\vec{q}_{\perp},\,-\vec{q}_{\perp}^{2}/2P^{+}\right)}{\left[-\frac{\vec{q}_{\perp}^{2}}{2P^{+}}-k^{-}-\frac{P^{-}}{2}+\frac{\left(\vec{k}_{\perp}-\vec{q}_{\perp}\right)^{2}+m^{2}}{2P^{+}\,x_{1}}-\frac{i\epsilon\,\left(1-x_{1}\right)}{2P^{+}\,x_{1}}\right]\left[-\frac{\vec{q}_{\perp}^{2}}{2P^{+}}+k^{\prime-}-\frac{P^{-}}{2}+\frac{\left(\vec{k}_{\perp}^{\prime}+\vec{q}_{\perp}\right)^{2}+m^{2}}{2P^{+}\,x_{2}}-\frac{i\epsilon\left(1-x_{2}\right)}{2P^{+}\,x_{2}}\right]}\right.\nonumber \\
&&-\frac{i\,\theta\left(-x_{1}\right)\,\theta\left(x_{2}\right)\,h\left(k,\,k^{\prime},\,\vec{q}_{\perp},\,k^{-}+\frac{P^{-}}{2}-\frac{\left(\vec{k}_{\perp}-\vec{q}_{\perp}\right)^{2}+m^{2}}{2P^{+}\,x_{1}}\right)}{\left[k^{-}+\frac{P^{-}}{2}-\frac{\left(\vec{k}_{\perp}-\vec{q}_{\perp}\right)^{2}+m^{2}}{2P^{+}\,x_{1}}\frac{\vec{q}_{\perp}^{2}}{2P^{+}}+\frac{i\epsilon\,\left(1-x_{1}\right)}{2P^{+}\,x_{1}}\right]\left[k^{-}+k^{\prime-}-\frac{\left(\vec{k}_{\perp}-\vec{q}_{\perp}\right)^{2}+m^{2}}{2P^{+}\,x_{1}}+\frac{\left(\vec{k}_{\perp}^{\prime}+\vec{q}_{\perp}\right)^{2}+m^{2}}{2P^{+}\,x_{2}}+\frac{i\epsilon\left(x_{2}-x_{1}\right)}{2P^{+}\,x_{1}\,x_{2}}\right]}\nonumber \\
&&\left.-\frac{i\,\theta\left(x_{1}\right)\,\theta\left(-x_{2}\right)\,h\left(k,\,k^{\prime},\,\vec{q}_{\perp},\,k^{-}+\frac{P^{-}}{2}-\frac{\left(\vec{k}_{\perp}-\vec{q}_{\perp}\right)^{2}+m^{2}}{2P^{+}\,x_{1}}\right)}{\left[k^{-}+k^{\prime-}-\frac{\left(\vec{k}_{\perp}-\vec{q}_{\perp}\right)^{2}+m^{2}}{2P^{+}\,x_{1}}+\frac{\left(\vec{k}_{\perp}^{\prime}+\vec{q}_{\perp}\right)^{2}+m^{2}}{2P^{+}\,x_{2}}+\frac{i\epsilon\left(x_{2}-x_{1}\right)}{2P^{+}\,x_{1}\,x_{2}}\right]\left[k^{\prime-}-\frac{P^{-}}{2}+\frac{\left(\vec{k}_{\perp}^{\prime}+\vec{q}_{\perp}\right)^{2}+m^{2}}{2P^{+}\,x_{2}}-\frac{\vec{q}_{\perp}^{2}}{2P^{+}}-\frac{i\epsilon\left(1-x_{2}\right)}{2P^{+}\,x_{2}}\right]}\right\} \, .\nonumber
%\\&& 
\end{eqnarray}
The first line of this expression corresponds to the pion pole contribution
and, therefore, it is the one that could be 
described by a "two GPD" contribution.
Actually, the fact that both $x_{1,2}$ are negative prevents us from this
simple interpretation. In fact, what happens is that one of the remaining
integrals, the one over $k^{-}$ or that over $k^{\prime-},$ vanishes, because all poles
are in the same half complex plane. In the other two contributions
present in Eq.~(\ref{complicated}), at least one of the $x_{1,2}$
is negative. In both cases one of the remaining integrals over $k^{-}$
or $k^{\prime-}$ vanishes by the same reasons than in the first case.

We want to emphasize that the vanishing of all diagrams of the type of
Fig.~\ref{Other_Contrib_Diagrams} as well as the diagram related to
the dressing of the non local vertex, depicted in Fig.~\ref{Fig_Dressed_vertex}
take place for all the different dPDFs. From a physical point of view, the
position of the pole in the lower or the upper half complex plane
corresponds to the two temporal contributions or, in other words,
the position of the pole tells us if we are dealing with a particle
or an antiparticle. The vanishing of all these integrals is related
to the fact that we can not close a loop with only particles or
only antiparticles. In all these diagrams, the fact that $q^{+}=0$,
which prevents  the presence of a particle
and an antiparticle in the same non local vertex, guarantees that they do not give any
additional contribution.

\section{
\label{III}
Test of factorization: an approximation to dPDFs in terms of GPDs}

Since dPDFs are experimentally basically unknown, their size and properties
are often inferred 
in terms of one-body quantities.
In Refs.~\cite{blok1} and~\cite{blok2}
it has been shown that, in a mean field approach, neglecting
correlations between the involved partons, the dPDF 
{ $F_{q \bar q}$} in momentum space factorizes
in the product of two GPDs at zero skewness. The validity
of this factorization has been analyzed in a number of model calculations
where it has been found to fail in general, in particular in
the valence region at the momentum scale associated to the model
\cite{bag,noi2,noi1,kase,plb,JHEP2016,Traini:2016jru,Rinaldi:2018zng}.
%In Ref.~\cite{Diehl1} it has been shown that the product of two GPDs
%is recovered by inserting a complete set of on-shell hadron states
%in between the two bilocal operators in the definition of the hadron dPDF
%in momentum space.
Let us now check whether or not the factorization in two GPDs, in the vector sector where it makes
sense for a pion,

\begin{equation}
F_{u\bar{d}}^{\left(2GPD\right)}\left(x_{1},x_{2},\vec{q}_{\perp}\right)=H_{\pi^{+}}^{u}\left(x_{1},0,t\right)\,H_{\pi^{+}}^{d}\left(-x_{2},0,t\right) \, ,
\label{2gpd}
\end{equation}
with $t=-\vec{q}_{\perp}^{\,2}$, is a good approximation
to the full result of the present NJL approach, given by Eq.~\eqref{A.16}.

To illustrate this property practically, we evaluate
integrals over the longitudinal variable
$x_2$ for both quantities, the dPDF  Eq.~\eqref{A.16}

\begin{equation}
    \label{int_dpdf}
    \bar F_{u\bar{d}}^{\left(0\right)}\left(x_{1},\vec{q}_{\perp}\right) 
    = \int d x_2 \,
    F_{u\bar{d}}^{\left(0\right)}\left(x_{1},x_{2},\vec{q}_{\perp}\right) 
\end{equation}
and the expression  Eq.~\eqref{2gpd} :

\begin{equation}
    \label{int_2gpd}
    \bar F_{u\bar{d}}^{\left(2GPD\right)}\left(x_{1},\vec{q}_{\perp}\right)
    = \int d x_2 \,
    F_{u\bar{d}}^{\left(2GPD\right)}\left(x_{1},x_{2},\vec{q}_{\perp}\right)
= H_{\pi^{+}}^{u}\left(x_{1},0,t\right) F_{\pi^+}(t)\,,
\end{equation}
where, in the last line, use has been made
of 
well known  sum rules for GPDs, with $F_{\pi^+}(t)$ the pion electromagnetic form factor.
\begin{figure}
\begin{centering}
\includegraphics[scale=0.42]{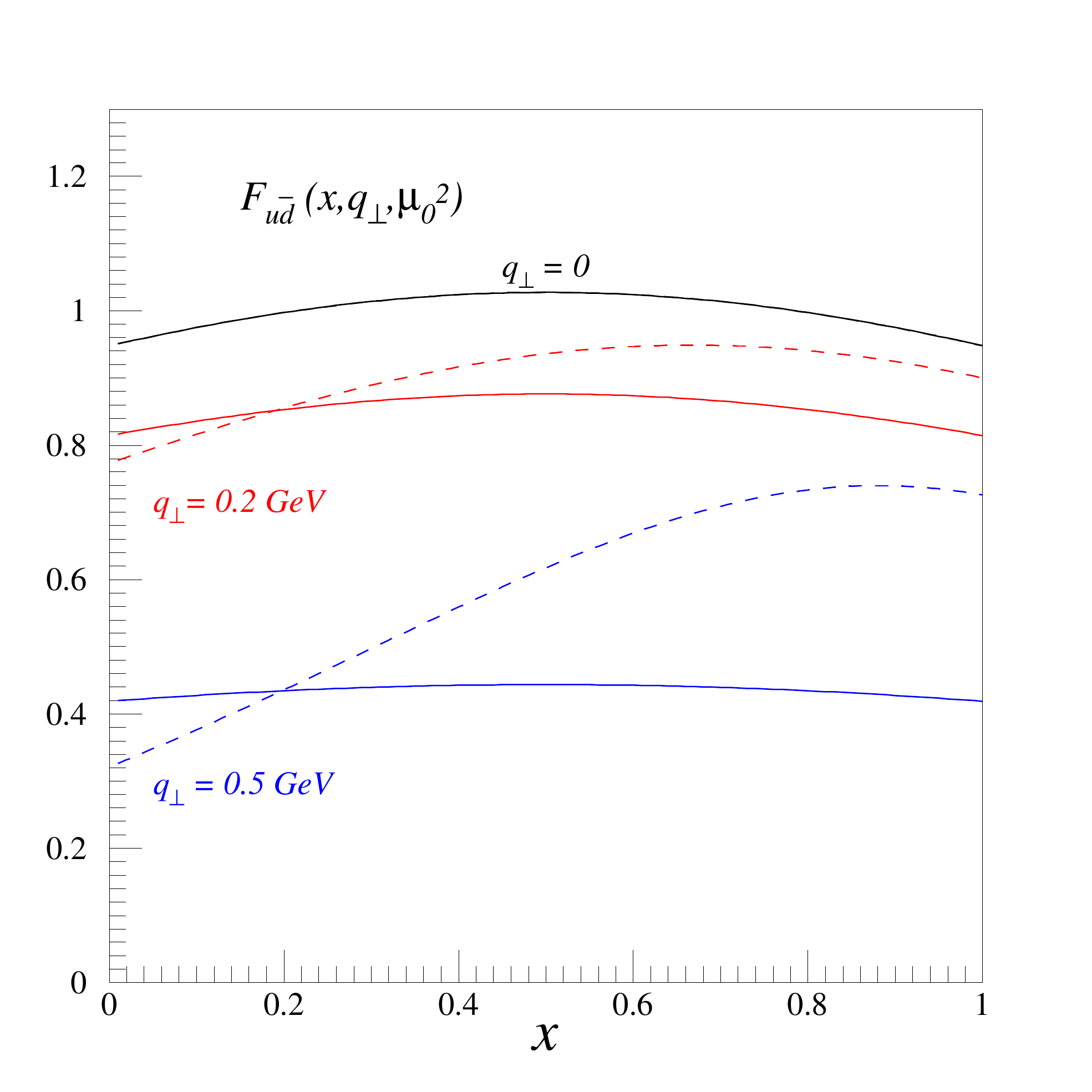}
\includegraphics[scale=0.67]{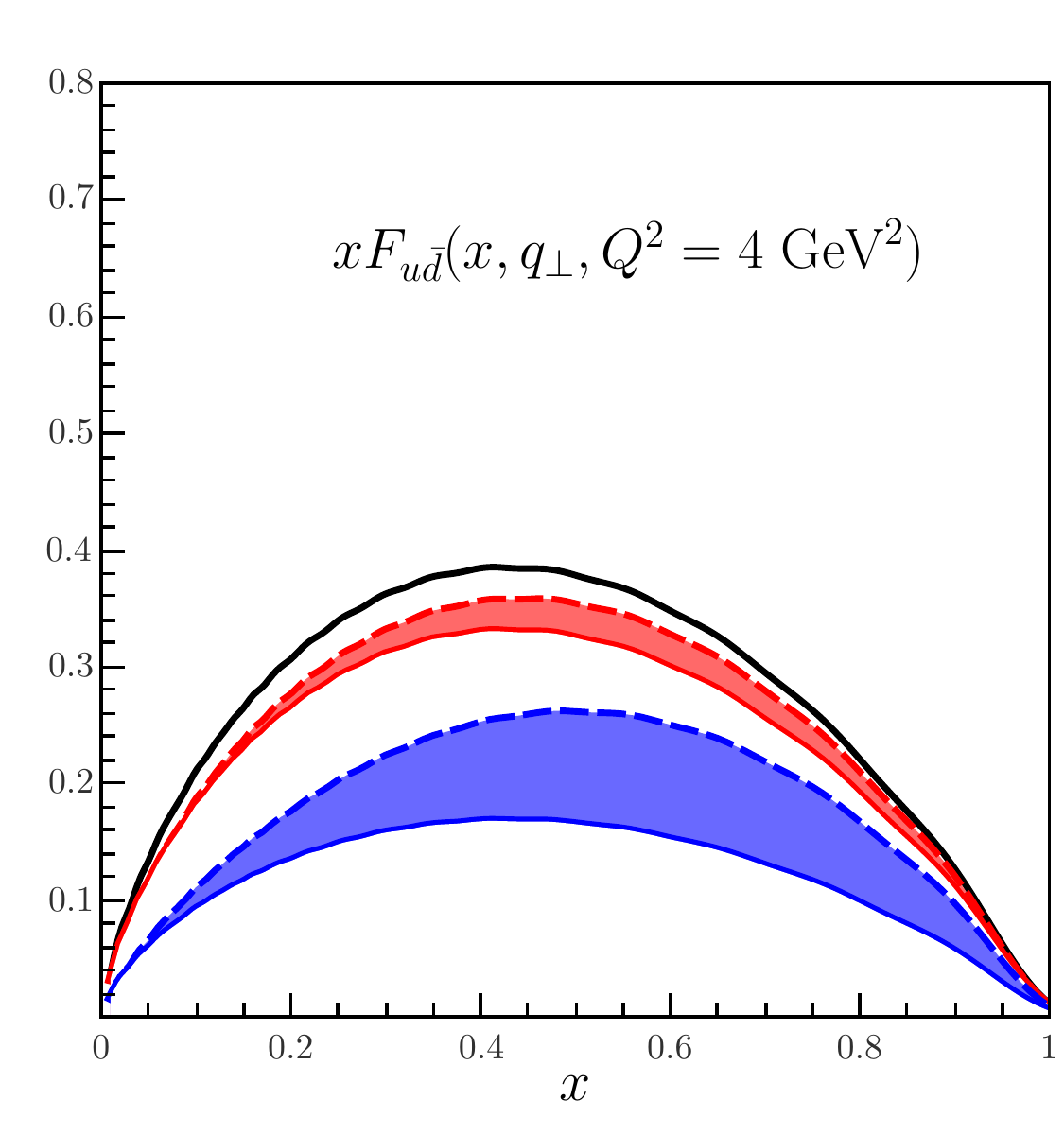}
\par\end{centering}
\caption{
Comparison between Eq.~\eqref{int_dpdf} (full) and its approximation Eq.~\eqref{int_2gpd} (dashed).
Left panel: results at the hadronic scale, $\mu_o$ = 0.29 GeV,
for $q_\perp=0,0.2,0.5$ GeV. Right panel: the same as in the previous panel,
but after LO QCD evolution to the momentum scale $Q^2$= 4 GeV$^2$ and multiplied, for an easy presentation, by the longitudinal momentum variable $x$. The  quality  of  the  approximation decreases as $q_\perp$ increases
as emphasized by the shaded areas, showing the
difference between the exact calculation and the approximation.}
\label{Fig2gpd}
\end{figure}
The comparison between Eqs.~\eqref{int_dpdf} and~\eqref{int_2gpd}
is shown in the left panel of Fig.~\ref{Fig2gpd}, using results of ref.~\cite{Theussl:2002xp}
for the NJL GPD, for three low values
of $q_\perp$, i.e., 0, 0.2 and 0.5 GeV.
It is clear that the approximation in terms of two GPDs holds
exactly at $q_\perp=0$, while it does not
work at higher values of $q_\perp$, in the present NJL framework, in the valence sector.
A similar conclusion was indeed obtained for the nucleon
in Refs.~\cite{bag,noi2,noi1,kase,plb,JHEP2016,Traini:2016jru},
using low energy models, and
recently, for the pion, using a light-front approach,
in Ref.~\cite{Rinaldi:2018zng}.

As already stated in the { first section}, 
the parton distribution obtained within a model should be associated to a low momentum scale, the so-called hadronic scale $\mu_0^2$.
For the pion in the NJL model such a value can be  fixed
in $\mu_0=$ 0.29 GeV (see, e.g. Ref. \cite{Courtoythesis}).
Since the approximation
Eq.~\eqref{2gpd} has been proposed for possible experimental observables
measured at colliders, such as the LHC, at high momentum scales and typically at low values of the longitudinal momentum fractions $x_1$ and $x_2$, it is important
to check whether the approximation works better
at high $Q^2$ values. We have therefore performed the leading order QCD evolution
of our results, from the scale $\mu_0^2$ to $Q^2=4$ GeV$^2$, following
the evolution procedure described in Ref.~\cite{Rinaldi:2018zng}.
The result is shown in the right panel of Fig.~\ref{Fig2gpd}.
It is clear that the difference between Eqs.~\eqref{int_dpdf} and~\eqref{int_2gpd}
persists in the present NJL scenario, at least in the valence sector, even at high momentum scales, as found in 
Ref.~\cite{Rinaldi:2018zng} with a different dynamical input.
Our model calculation shows that relevant novel information on two-body parton correlations, that are  not included in one body quantities nor described in a mean field approach, would be 
accessible through experimental or lattice measurements
of dPDFs.

%\[
%--------------------------------
%\]

%{\bf TEXT REMOVED (?)}

%\begin{figure}[htbp]
%\begin{center}
%\includegraphics[scale=1.0]{Fig_yt2.pdf}
%\caption{plot}
%\label{Fig:6}
%\end{center}
%\end{figure}

%{\it Forma alternativa (que no mejor) de notacion: 
%\begin{align*}
%j\left(P,q,x\right) & =\theta\left(x\right)\theta\left(1-x\right)\frac{1}{4\pi^{2}}\sum_{j=0}^{2}c_{j}\left[-\ln\frac{\mu_{j}\left(x\right)^{2}}{\mu_{0}\left(x\right)^{2}}-\right.\\
% & \left.\frac{A-2m_{\pi}^{2}x\left(1-x\right)}{\sqrt{A^{2}-4\,\mu_{j}\left(x\right)^{2}\,q^{2}\left(1-x\right)^{2}}}\text{arccoth}\frac{A+\mu_{j}\left(x\right)^{2}}{\sqrt{A^{2}-4\,\mu_{j}\left(x\right)^{2}\,q^{2}\left(1-x\right)^{2}}}-\frac{1}{2}\ln\frac{B+\mu_{j}\left(x\right)^{2}}{\mu_{j}\left(x\right)^{2}}\right]
%\end{align*}
%where 
%\begin{align*}
%\mu_{j}\left(x\right)^{2} & =M_{j}^{2}-m_{\pi}^{2}x\left(1-x\right)\\
%A & =\left(-q^{2}+2P\cdot q\,x\right)\left(1-x\right)\\
%B & =\left(-q^{2}+2P\cdot q\right)\,x\left(1-x\right)
%\end{align*}
%En este caso habria que volver hacia atras e introducir $\mu_{j}\left(x\right)$ desde (\ref{A.16}).}

%\[
%--------------------------------
%\]

\section{\label{IV} Conclusions}

A consistent field-theoretical approach, based
on the  Nambu--Jona-Lasinio  model, 
with
two different regulariztion schemes, the standard Pauli-Villars method and
a properly introduced Light-Front
one, is used for
a systematic analysis 
of double parton distribution functions in the pion.
Results are presented for several 
double parton distributions corresponding to different Dirac operators in their definitions.
In particular, in the vector sector,
it is found that
these functions encode novel non-perturbative information,
not present in one-body quantities, such as PDFs and GPDs, as it happens in model calculations
of proton dPDFs as well as in the only phenomenological
evaluation of pion dPDFs available at present.
{ In particular, we have shown that the approximation 
of the momentum space spin-independent dPDF in terms of two GPDs at zero skewness does not hold in our approach.} 
This fact is true also after QCD evolution of the model results, {the latter being} associated
to a low hadronic scale, to experimental
high momentum scales.

Lattice data have been already
obtained for two current correlations in the pion,
quantities related to dPDFs.
The analysis of two curent correlations in the pion in the NJL
would correspond to a completely different calculation, presently in progress \cite{inprog},
with respect to the one presented here.
The evaluation of pion dPDFs on the lattice has been planned
\cite{private}.
It will be interesting to compare our results
for the physical dPDFs with the forth-coming lattice data.

\section*{Acknowledgments}

We thank M. Rinaldi for useful discussions. 
This work was supported in part by the Mineco
under contract FPA2016-77177-C2-1-P, by
the Centro de Excelencia Severo Ochoa Programme grant SEV-2014-0398, {by the 
STRONG-2020 project of the European Union’s Horizon 2020 research and innovation programme 
under grant agreement No 824093,}  and by UNAM through the PIIF project Perspectivas en F\'isica de Part\'iculas y Astropart\'iculas as well as  Grant No. DGAPA-PAPIIT IA102418.
A.C. and S.S. thank the Department of Theoretical Physics of the
University of Valencia for warm hospitality and support; S.N. thanks the INFN,
sezione di Perugia,  
and the Department of Physics and Geology of the University
of Perugia for warm hospitality and support.

\appendix

\section{The NJL model and regularization scheme}

\label{App.NJL_regularization}

\subsection{Basic physical quantities in the NJL model}
\label{App.NJL_Basic_Equations}

The Lagrangian density in the two-flavor version of the NJL model
is \cite{Klevansky:1992qe} 
\begin{equation}
\mathcal{L}=\bar{\psi}\left(i\,\not\partial-m_{0}\right)\psi+g\left[\left(\bar{\psi}\,\psi\right)^{2}+\left(\bar{\psi}\,\vec{\tau}\,i\gamma_{5}\,\psi\right)^{2}\right]\,\,,
\end{equation}
where $m_{0}$ is the current quark mass. The NJL is a chiral theory
that reproduces the spontaneous symmetry breaking process in which
the quark mass moves from the current value to its constituent value,
\begin{equation}
m=m_{0}-4g\,\left\langle \bar{u}u\right\rangle \,,
\end{equation}
where $\left\langle \bar{u}u\right\rangle $ is the quark condensate.

The main physical quantities associated to pion physics are defined
in terms of two integrals: 
\begin{equation}
I_{1}\left(m\right)=i\int\frac{d^{4}k}{\left(2\pi\right)^{4}}\frac{1}{k^{2}-m^{2}+i\epsilon}\label{I1.1}
\end{equation}
\begin{equation}
I_{2}\left(m,q^{2}\right)=i\int\frac{d^{4}k}{\left(2\pi\right)^{4}}\frac{1}{\left[\left(k+\frac{q}{2}\right)^{2}-m^{2}+i\epsilon\right]\left[\left(k-\frac{q}{2}\right)^{2}-m^{2}+i\epsilon\right]}\label{I2.1}
\end{equation}

Effectively, in the large $N_{c}$ approximation, the quark condensate
is defined by
\begin{equation}
\left\langle \bar{u}u\right\rangle =-4N_{c}mI_{1}\,.
\end{equation}
Pion and sigma masses are defined by the relations 
\begin{equation}
2g\,\Pi_{PS}\left(m_{\pi}^{2}\right)=1\,\,\,\,\,\,\,\,\,,\,\,\,\,\,\,\,\,\,\,\,\,\,2g\,\Pi_{S}\left(m_{\sigma}^{2}\right)=1\,,\label{NJL.17}
\end{equation}
with the scalar polarization 
\begin{align}
\Pi_{S}\left(q^{2}\right) & =-i\int\frac{d^{4}k}{\left(2\pi\right)^{4}}\text{Tr}\left[iS_{F}\left(p\right)\,iS_{F}\left(p-q\right)\right]\nonumber \\
 & =8N_{c}\left[I_{1}+\frac{1}{2}\left(4m^{2}-q^{2}\right)I_{2}\left(q\right)\right]\,,\label{NJL.14}
\end{align}
and the pseudoscalar polarization 
\begin{align}
\Pi_{PS}\left(q^{2}\right) & =-i\int\frac{d^{4}k}{\left(2\pi\right)^{4}}\text{Tr}\left[i\gamma_{5}\tau^{i}\,iS_{F}\left(p\right)i\gamma_{5}\tau^{i}\,iS_{F}\left(p-q\right)\right]\nonumber \\
 & =8N_{c}\left[I_{1}-\frac{1}{2}q^{2}I_{2}\left(q\right)\right]\,.\label{NJL.15}
\end{align}
The pion-quark and sigma-quark coupling constant are respectively
defined by 
\begin{align}
g_{\pi qq}^{2} & =\left(\frac{\partial\Pi_{PS}\left(q^{2}\right)}{\partial q^{2}}\right)_{q^{2}=m_{\pi}^{2}}^{-1}=\frac{-1}{4N_{c}\left[I_{2}\left(m_{\pi}^{2}\right)+m_{\pi}^{2}\left(\partial I_{2}/\partial q^{2}\right)_{q^{2}=m_{\pi}^{2}}\right]}\,,\nonumber \\
\label{NJL.20}\\
g_{\sigma qq}^{2} & =\left(\frac{\partial\Pi_{S}\left(q^{2}\right)}{\partial q^{2}}\right)_{q^{2}=m_{\sigma}^{2}}^{-1}=\frac{-1}{4N_{c}\left[I_{2}\left(m_{\sigma}^{2}\right)-\left(4m^{2}-m_{\sigma}^{2}\right)\left(\partial I_{2}/\partial q^{2}\right)_{q^{2}=m_{\sigma}^{2}}\right]}\,.\nonumber 
\end{align}
The pion decay constant are
\begin{equation}
f_{\pi}=-4N_{c}g_{\pi qq}mI_{2}\left(m_{\pi}^{2}\right)\,,
\end{equation}

The NJL model is a non-renormalizable field theory and a regularization
procedure has to be defined for the calculation of $I_{1}\left(m\right)$
and $I_{2}\left(m,q^{2}\right).$ We will introduce now the Pauli-Villars
regularization method for the NJL model and, in section \ref{App.NJL_LF_regularization},
we built a regularization method adapted for calculations in the Light
Front formalism.

\subsection{Pauli Villars regularization scheme}
\label{App.NJL_PV_regularization}

In Section \ref{II}, we have used the Pauli-Villars regularization
in order to render the occurring integrals finite. The way to proceed
in this method is: 
(1) remove from the numerator all the powers of the integrated momentum, which will be replaced by external momenta, and the mass of the constituent
quark, $m$; 
(2) for each resulting integral, which is of the form
\begin{equation}
\widetilde{I}_{n}\left(\mu\left(m\right)\right)=\int\frac{d^{4}k}{\left(2\pi\right)^{4}}\frac{1}{\left[k^{2}-\mu\left(m\right)^{2}+i\epsilon\right]^{n}}\,,
\end{equation}
make the substitution
\begin{equation}
\widetilde{I}_{n}^{r}\left(\mu\left(m\right)\right)=\sum_{j=0}^{2}\,c_{j}\,\widetilde{I}_{n}\left(\mu\left(M_{j}\right)\right)\,,
\end{equation}
with $M_{j}^{2}=m^{2}+j\,\Lambda^{2}$, $c_{0}=c_{2}=1$ and $c_{1}=-2$.

Following this procedure, we obtain for the momentum integral of one propagator
\begin{equation}
I_{1}=\frac{1}{16\pi^{2}}\sum_{j=0}^{2}c_{j}M_{j}^{2}\ln\frac{M_{j}^{2}}{m^{2}}\,,
\end{equation}
and for the one of two propagators  
\begin{align}
I_{2}\left(m,q^{2}\right) & =\frac{1}{16\pi^{2}}\sum_{j=0}^{2}c_{j}\left(\ln\frac{M_{j}^{2}}{m^{2}}+2\sqrt{\frac{4M_{j}^{2}}{q^{2}}-1}\,\arctan\frac{1}{\sqrt{\frac{4M_{j}^{2}}{q^{2}}-1}}\right)\,.
\end{align}
With the conventional values $\left\langle \bar{u}u\right\rangle =-(0.250\,\text{GeV})^{3},$
$f_{\pi}=0.0924\,\text{GeV}$ and $m_{\pi}=0.140\,\text{GeV}$, we
get $m=0.238\,\text{GeV}$, $\Lambda$=0.860 GeV and $m_{0}=5.4\,\text{MeV}.$
For the pion-quark coupling constant we get $g_{\pi qq}^{2}=6.279.$
We can obtain the chiral limit taking $m_{0}=0$, without changing
$\Lambda$ and $m.$ In that case $\left\langle \bar{u}u\right\rangle $
and $f_{\pi}$ do not change but one has $m_{\pi}=0$ and $g_{\pi qq}^{2}=6.625.$

\subsection{A regularization of the NJL model in the Light Front}
\label{App.NJL_LF_regularization}

The Pauli-Villars (PV) regularization scheme, which respects
the gauge symmetry of the problem, has been often adopted. Nevertheless,
this procedure is based on an equal time quantization of the field
theory, while the dPDF are defined in the light front formalism. In many cases
this point has no consequences, but sometimes it does, as we will
see later. Other usual regularization schemes, like the covariant
four-momentum cutoff, the three momentum cutoff or the proper time
regularization, are also defined in the equal time quantization of
the field theory and they are manifestly not useful in a light front
calculation.

Our aim, in this section, is to define a regularization procedure
for the NJL model which respects the light front formalism. The ideal
scheme will be: (1) to integrate $k^{-}$ using the poles of the propagators and,
as a result of this integration, the range of variation of $k^{+}$
will be bounded; (2) introducing a cut-off,$\left|\vec{k}_{\perp}\right|<\Lambda_{\perp}$,
to perform the integration over $\vec{k}_{\perp}$ and the integration
over the bounded range of variation of $k^{+}$.

To have a clear notation, we call $\tilde{m}_{0}$ and $\tilde{m}$ the current and constituent quark masses evaluated in this aproach. The gap equation,
\begin{equation}
\tilde{m}=\tilde{m}_{0}-4g\,\left\langle \bar{u}u\right\rangle \,,
\end{equation}
will be always valid.

The defined procedure works for $I_{2}\left(\tilde{m},q^{2}\right).$ Effectively,
after integration over $k^{-}$ and introducing the change of variable
$k^{+}=\left(x-\frac{1}{2}\right)\left|q^{+}\right|$, we have from
Eq.~(\ref{I2.1}) (for simplicity we can choose $q^{\mu}=\left(q^{+},\vec{0}_{\perp},q^{-}\right)$
and $q^{2}=2\,q^{+}q^{-}),$
\begin{equation}
I_{2}\left(\tilde{m},q^{2}\right)=-\frac{1}{16\,\pi^{2}}\int_{0}^{1}\,dx\,\int_{0}^{\Lambda_{\perp}^{2}}\,dk_{\perp}^{2}\,\frac{1}{k_{\perp}^{2}+\tilde{m}^{2}-q^{2}x\left(1-x)\right)}\,,\label{I2.2}
\end{equation}
and, performing these two integrations we arrive to
\begin{equation}
I_{2}\left(\tilde{m},q^{2}\right)=\frac{1}{16\,\pi^{2}}\left[\ln\frac{\tilde{m}^{2}}{\Lambda_{\perp}^{2}+\tilde{m}^{2}}+\phi\left(\frac{q^{2}}{\tilde{m}^{2}}\right)-\phi\left(\frac{q^{2}}{\Lambda_{\perp}^{2}+\tilde{m}^{2}}\right)\right]\quad,
\label{I2.3a}
\end{equation}
with
\begin{equation}
\phi\left(z\right)=\begin{cases}
\sqrt{1-4/z}\,\ln\frac{\sqrt{1-4/z}+1}{\sqrt{1-4/z}-1}\,, & z<0\\
\\
2\sqrt{\frac{4}{z}-1}\arctan\frac{1}{\sqrt{\frac{4}{z}-1}}\,, & 0<z<4\\
\\
\sqrt{1-4/z}\,\left(\ln\frac{1+\sqrt{1-4/z}}{1-\sqrt{1-4/z}}-i\pi\right)\,, & 4<z\quad.
\end{cases}
\end{equation}
%\textcolor{red}{(opción 2, compacta)}
%\begin{align}
%I_{2}\left(\tilde{m},q^{2}\right) & =\frac{1}{16\,\pi^{2}}\left[\ln\frac{\tilde{m}^{2}}{\Lambda_{\perp}^{2}+\tilde{m}^{2}}+2\sqrt{\frac{4\,\tilde{m}^{2}}{q^{2}}-1}\arctan\frac{1}{\sqrt{\frac{4\,\tilde{m}^{2}}{q^{2}}-1}}\right.\nonumber \\
% & \left.\,\,\,\,\,\,\,\,\,\,\,\,\,\,\,\,\,\,\,\,\,\,\,\,-2\sqrt{\frac{4\,\left(\Lambda_{\perp}^{2}+\tilde{m}^{2}\right)}{q^{2}}-1}\arctan\frac{1}{\sqrt{\frac{4\,\left(\Lambda_{\perp}^{2}+\tilde{m}^{2}\right)}{q^{2}}-1}}\right]\,.\label{I2.3b}
%\end{align}

The procedure that has allowed us to evaluate $I_{2}\left(\tilde{m},q^{2}\right)$
is not sufficient to determine $I_{1}\left(\tilde{m}\right).$ In fact, this
is a well known problem: $I_{1}\left(\tilde{m}\right)$ corresponds to a tadpole
diagram associated to the vacuum condensate, which is evaluable in
a regularization scheme in an equal time formulation of the quantum
field theory (like PV regulaization), but needs for additional assumptions
in order to be evaluated in a light front formulation \cite{Burkardt:1995ct,Burkardt:1997bd}.

Nevertheless, some information on $I_{1}\left({m}\right)$ can be obtained
from $I_{2}\left({m},q^{2}\right)$,
using the fact that 
\begin{equation}
\frac{d}{d\tilde{m}^{2}}I_{1}\left(\tilde{m}\right)=\lim_{q^{\mu}\rightarrow0}\,I_{2}\left(\tilde{m},q^{2}\right)\,.\label{DI1_I2}
\end{equation}
From our result Eq.~\eqref{I2.3a}
we have 
\begin{equation}
I_{1}\left(\tilde{m}\right)=\frac{1}{16\,\pi^{2}}\left[C\left(\Lambda_{\perp}\right)-\left(\Lambda_{\perp}^{2}+\tilde{m}^{2}\right)\ln\frac{\Lambda_{\perp}^{2}+\tilde{m}^{2}}{\Lambda_{\perp}^{2}}+\tilde{m}^{2}\ln\frac{\tilde{m}^{2}}{\Lambda_{\perp}^{2}}\right]\,,\label{I1.2}
\end{equation}
where $C\left(\Lambda_{\perp}\right)$ is an arbitrary function of
$\Lambda_{\perp}.$ Therefore, Eq. (\ref{DI1_I2}) fixes the dependence of
 $I_{1}\left(\tilde{m}\right)$ 
on $\tilde{m}$,
but we need some additional input
to fix its dependence on $\Lambda_{\perp}.$

We can perform explicitly the $k^{-}$ integral present in $I_{1}\left(\tilde{m}\right)$
obtaining 
\begin{equation}
I_{1}\left(\tilde{m}\right)=\int\frac{d^{2}k_{\perp}}{16\,\pi^{3}}\int_{0}^{\infty}\frac{dk^{+}}{k^{+}}\,.
\end{equation}
Now, for the $k^{+}$ integral we introduce an infrared and an ultraviolet
cut-off imposing \cite{Itakura:2000te} 
\begin{equation}
\frac{k_{\perp}^{2}+\tilde{m}^{2}}{\Lambda^{+}}<k^{+}<\Lambda^{+}\,,\label{I1.4}
\end{equation}
and we arrive to 
\begin{eqnarray}
I_{1}\left(\tilde{m}\right) & = & \int_{0}^{\Lambda_{\perp}^{2}}\frac{dk_{\perp}^{2}}{16\,\pi^{2}}\ln\frac{2\Lambda^{+2}}{k_{\perp}^{2}+\tilde{m}^{2}}\nonumber \\
 & = & \frac{1}{16\,\pi^{2}}\left[\Lambda_{\perp}^{2}\left(1+\ln\frac{2\,\Lambda^{+2}}{\Lambda_{\perp}^{2}}\right)-\left(\Lambda_{\perp}^{2}+\tilde{m}^{2}\right)\ln\frac{\Lambda_{\perp}^{2}+\tilde{m}^{2}}{\Lambda_{\perp}^{2}}+\tilde{m}^{2}\ln\frac{\tilde{m}^{2}}{\Lambda_{\perp}^{2}}\right]\,. \label{I1.6}
\end{eqnarray}
This result is consistent with Eq.~(\ref{I1.2}) and, in agreement
with the discussion in
Ref. \cite{Burkardt:1995ct,Burkardt:1997bd}, we need an additional
information because we have two different cutoffs. On the mass shell
we have that $k^{+}=\frac{1}{\sqrt{2}}\left(E+k^{3}\right)$ and,
assuming that $k^{3}<\Lambda_{\perp},$ a natural way to relate both
cutoffs is to use $\Lambda^{+}=\sqrt{2}\Lambda_{\perp}.$ In this
way, we finally obtain 
\begin{equation}
I_{1}\left(\tilde{m}\right)=\frac{1}{16\,\pi^{2}}\left[\Lambda_{\perp}^{2}\left(1+\ln4\right)-\left(\Lambda_{\perp}^{2}+\tilde{m}^{2}\right)\ln\frac{\Lambda_{\perp}^{2}+\tilde{m}^{2}}{\Lambda_{\perp}^{2}}+\tilde{m}^{2}\ln\frac{\tilde{m}^{2}}{\Lambda_{\perp}^{2}}\right]
\,.
\label{I1.7}
\end{equation}

As explained in Ref. \cite{Itakura:2000te}, the choice made in Eq. (\ref{I1.4})
for the cutoffs, relating the ultraviolet and the infrared ones, is
natural on the basis of the respect of the dispersion relation $2k^{+}k^{-}-k_{\perp}^{2}=\tilde{m}^{2}$
and on the restoration of the symmetry between $k^{+}$ and $k^{-}.$
Parity transformation implies the exchange of $k^{+}$ and $k^{-},$
therefore the minimum value of $k^{+}$ must be related to the maximum
value of $k^{-}$ through the dispersion relation and the assumption
of the same maximum value for $k^{+}$ and $k^{-}.$

In this scheme,
with the conventional values $\left\langle \bar{u}u\right\rangle =-(0.250\,\text{GeV})^{3},$
$f_{\pi}=0.0924\,\text{GeV}$ and $m_{\pi}=0.140\,\text{GeV}$, we
get $\tilde{m}=0.246\,\text{GeV}$, $\Lambda_{\perp}$=0.572 GeV and $\tilde{m}_{0}=5.3\,\text{MeV}.$
For the pion-quark coupling constant we get $\tilde{g}_{\pi qq}^{2}=6.735.$
We can obtain the chiral limit taking $\tilde{m}_{0}=0$, without changing
$\Lambda_{\perp}$ and $\tilde{m}.$ In that case $\left\langle \bar{u}u\right\rangle $
and $f_{\pi}$ do not change but one has $m_{\pi}=0$ and $\tilde{g}_{\pi qq}^{2}=7.085\,.$

\subsection{Some intermediate results}
\label{App.Intermediate-Results}

The traces involved in Eq. \eqref{A.11} are: 
\begin{equation}
\begin{array}{lll}
tr_{u,\bar{d}}\left(\vec{k}_{\perp},\,\vec{q}_{\perp}\right) & = & -2\,P^{+2}\,\left[\vec{k}_{\perp}^{\thinspace2}-\frac{1}{4}\vec{q}_{\perp}^{\thinspace2}+m^{2}\right]\\
tr_{\Delta u,\Delta\bar{d}}\left(\vec{k}_{\perp},\,\vec{q}_{\perp}\right) & = & -2\,P^{+2}\,\left[\vec{k}_{\perp}^{\thinspace2}-\frac{1}{4}\vec{q}_{\perp}^{\thinspace2}-m^{2}\right]\\
tr_{\delta u^{j},\delta\bar{d}^{k}}\left(\vec{k}_{\perp},\,\vec{q}_{\perp}\right) & = & 2\,P^{+2}\,\left[-2\,k_{\perp}^{j}\,k_{\perp}^{k}+\frac{1}{2}\,q_{\perp}^{j}\,q_{\perp}^{k}+\delta^{j,k}\,\left(\vec{k}_{\perp}^{\thinspace2}-\frac{1}{4}\vec{q}_{\perp}^{\thinspace2}+m^{2}\right)\right]\\
tr_{S,S}\left(\vec{k}_{\perp},\,\vec{q}_{\perp}\right) & = & \frac{1}{2}m_{\pi}^{2}\,\left(4\,m^{2}+\vec{q}_{\perp}^{\thinspace2}\right)\\
tr_{P,P}\left(\vec{k}_{\perp},\,\vec{q}_{\perp}\right) & = & \frac{1}{2}m_{\pi}^{2}\,\vec{q}_{\perp}^{\thinspace2}
\end{array}
\end{equation}
\begin{equation}
\begin{array}{llccl}
tr_{u,\Delta\bar{d}}\left(\vec{k}_{\perp},\,\vec{q}_{\perp}\right) & = & tr_{\Delta u,\bar{d}}\left(\vec{t}_{\perp},\,\vec{q}_{\perp},y\right) & = & -2\,i\,P^{+2}\,\epsilon^{j,k}\,k_{\perp}^{j}\,q_{\perp}^{k}\\
tr_{u,\delta\bar{d}^{j}}\left(\vec{k}_{\perp},\,\vec{q}_{\perp}\right) & = & -tr_{\delta u^{j},\bar{d}}\left(\vec{t}_{\perp},\,\vec{q}_{\perp}\right) & = & 2\,i\,P^{+2}\,m\,\epsilon^{j,k}\,q_{\perp}^{k}\\
tr_{\Delta u,\delta\bar{d}^{j}}\left(\vec{k}_{\perp},\,\vec{q}_{\perp}\right) & = & -tr_{\delta u^{j},\Delta\bar{d}}\left(\vec{t}_{\perp},\,\vec{q}_{\perp}\right) & = & 4\,m\,P^{+2}\,k_{\perp}^{j}
\end{array}
\end{equation}
with $\epsilon^{j,k}=\epsilon^{0,j,k,3}.$
 
Using the relation
\begin{align}
\int\frac{d^{2}k_{\perp}}{\left(2\pi\right)^{2}} & \frac{k_{\perp}^{j}\,k_{\perp}^{k}}{\left[\left(\vec{k}_{\perp}+\frac{q_{\perp}}{2}\right)^{2}+m^{2}-x_{1}\left(1-x_{1}\right)m_{\pi}^{2}-i\epsilon\right]\left[\left(\vec{k}_{\perp}-\frac{q_{\perp}}{2}\right)^{2}+m^{2}-x_{1}\left(1-x_{1}\right)m_{\pi}^{2}-i\epsilon\right]}=\nonumber \\
\int\frac{d^{2}k_{\perp}}{\left(2\pi\right)^{2}} & \frac{\left[\delta^{jk}\left(k_{\perp}^{2}-\frac{1}{q_{\perp}^{2}}\left(\vec{k}_{\perp}\cdot\vec{q}_{\perp}\right)^{2}\right)+\hat{q}_{\perp}^{j}\,\hat{q}_{\perp}^{k}\left(\frac{2}{q_{\perp}^{2}}\left(\vec{k}_{\perp}\cdot\vec{q}_{\perp}\right)^{2}-k_{\perp}^{2}\right)\right]}{\left[\left(\vec{k}_{\perp}+\frac{q_{\perp}}{2}\right)^{2}+m^{2}-x_{1}\left(1-x_{1}\right)m_{\pi}^{2}-i\epsilon\right]\left[\left(\vec{k}_{\perp}-\frac{q_{\perp}}{2}\right)^{2}+m^{2}-x_{1}\left(1-x_{1}\right)m_{\pi}^{2}-i\epsilon\right]}\,
\end{align}
we observe that the tensor trace can be rewritten as
\begin{equation}
tr_{\delta u^{j},\delta\bar{d}^{k}}\left(\vec{k}_{\perp},\,\vec{q}_{\perp}\right)=2\,P^{+2}\,\left[\left(2\,\hat{q}_{\perp}^{j}\,\hat{q}_{\perp}^{k}-\delta^{j,k}\right)\left(k_{\perp}^{\thinspace2}+\frac{q_{\perp}^{\thinspace2}}{4}-\frac{2}{q_{\perp}^{2}}\left(\vec{k}_{\perp}\cdot\vec{q}_{\perp}\right)^{2}\right)+\delta^{j,k}\,m^{2}\right]\,,
\end{equation}
showing in an explicit form the tensor structure defined in Eq. (\ref{TensorStr_q})
for $F_{\delta q^{j},\delta\bar{q}^{k}}\left(x_{1},x_{2},\vec{q}_{\perp}^{\,2}\right).$

All these traces are of the generic form
\begin{equation}
tr_{a_{1}\bar{a}_{2}}\left(\vec{k}_{\perp},\,\vec{q}_{\perp}\right)=2\,P^{+2}\,\left[A\,k_{\perp}^{2}+\vec{B}\cdot\vec{k}_{\perp}+D+G\left(\vec{q}_{\perp}\cdot\vec{k}_{\perp}\right)^{2}\right]\,, \label{Ap.4.10}
\end{equation}
where $A,\,\vec{B},\,D$ and $G$ are functions of $\vec{q}_{\perp}$
and $m$ but $\vec{k}_{\perp}$-independent. The linear terms in $\vec{k}_{\perp}$
in Eq. (\ref{Ap.4.10}) will vanish after the $\vec{k}_{\perp}$ integration
present in Eq. (\ref{A.10}). Therefore, the final result can be written
as (we follow here the notation used in Eqs. (\ref{A.16}-\ref{at})),
\begin{equation}
F_{a_{1},\bar{a_{2}}}\left(x_{1},x_{2},\vec{q}_{\perp}\right)=-C\left(x_{1},x_{2}\right)\left[A\,I_{A}^{r}+\frac{D}{q_{\perp}^{2}}\,I_{D}^{r}+G\,q_{\perp}^{2}\,I_{G}^{r}\right]\,.
\end{equation}

In order to obtain Eq (\ref{A.16}-\ref{at}) in the Pauli-Villars
regularization scheme from Eq. (\ref{A.10}) we join the two propagators
using Feynman parametrization, we make the change of variable $\vec{k}_{\perp}=\vec{t}_{\perp}-\vec{q}_{\perp}\left(y-1/2\right)$
and we remove the $t_{\perp}^{2}$ present in the numerator before
integration. For instance, in the case of $I_{A}^{r}$ we have, before the regularization,
\begin{align}
I_{A} & =4\pi\,\int\frac{d^{2}k_{\perp}}{\left(2\pi\right)^{2}}\nonumber \\
 & \frac{k_{\perp}^{2}}{\left[\left(\vec{k}_{\perp}+\frac{q_{\perp}}{2}\right)^{2}+m^{2}-x_{1}\left(1-x_{1}\right)m_{\pi}^{2}-i\epsilon\right]\left[\left(\vec{k}_{\perp}-\frac{q_{\perp}}{2}\right)^{2}+m^{2}-x_{1}\left(1-x_{1}\right)m_{\pi}^{2}-i\epsilon\right]}\nonumber \\
 & =4\pi\,\left[\int_{0}^{1}\,dy\,\int\frac{d^{2}t_{\perp}}{\left(2\pi\right)^{2}}\frac{1}{\left[t_{\perp}^{2}+q_{\perp}^{2}y\left(1-y\right)+\kappa-i\epsilon\right]}\right.\nonumber \\
 & -2q_{\perp}^{2}\,\int_{0}^{1}\,dy\,\int\frac{d^{2}t_{\perp}}{\left(2\pi\right)^{2}}\frac{y\left(1-y\right)}{\left[t_{\perp}^{2}+q_{\perp}^{2}y\left(1-y\right)+\kappa-i\epsilon\right]^{2}}\nonumber \\
 & -\left.\left(\kappa-\frac{1}{4}q_{\perp}^{2}\right)\,\int_{0}^{1}\,dy\,\int\frac{d^{2}t_{\perp}}{\left(2\pi\right)^{2}}\frac{1}{\left[t_{\perp}^{2}+q_{\perp}^{2}y\left(1-y\right)+\kappa-i\epsilon\right]^{2}}\right]\,.
\end{align}
Now we perform each integral using standard methods and we regularize
it, obtaining
\begin{equation}
I_{A}^{r\left(PV\right)}=\sum_{i=0}^{2}\,c_{i}\,\left(-\ln\frac{\kappa_{i}}{\kappa}-\left(2\frac{\kappa}{q_{\perp}^{2}}+\frac{1}{2}\right)f\left(\frac{\kappa_{i}}{q_{\perp}^{2}}\right)\right)\,.
\end{equation}
Following the same procedure we have 
\begin{align}
I_{D}^{r\left(PV\right)} & =\sum_{i=0}^{2}\,c_{i}\,f\left(\frac{\kappa_{i}}{q_{\perp}^{2}}\right)\,,\\
I_{G}^{r\left(PV\right)} & =\sum_{i=0}^{2}\,c_{i}\,\left(-\frac{1}{2}\ln\frac{\kappa_{i}}{\kappa}-\frac{m^{2}-M_{i}^{2}}{q_{\perp}^{2}}\,f\left(\frac{\kappa_{i}}{q_{\perp}^{2}}\right)\right)\,.\nonumber 
\end{align}

Eqs. (\ref{A.16-1}-\ref{at-1}) have been obtained in the LF regularization
scheme. In this case we have, for the $I_{A}^{r}$ contribution,
\begin{equation}
I_{A}^{r\left(LF\right)}=4\pi\,\int_{0}^{\Lambda_{\perp}}\frac{dk_{\perp}}{\left(2\pi\right)^{2}}\,k_{\perp}\,\int_{0}^{2\pi}\,d\theta\,\frac{k_{\perp}^{2}}{\left(k_{\perp}^{2}+\frac{q_{\perp}^{2}}{4}+\tilde{\kappa}\right)^{2}-k_{\perp}^{2}\,q_{\perp}^{2}\,\cos^{2}\theta}\,.
\end{equation}
Here, both integrals, on $\theta$ and on $k_{\perp},$ can be performed
obtaining
\begin{equation}
I_{A}^{r\left(LF\right)}=g_{2}\left(q_{\perp}\right)\,,
\end{equation}
and, in a similar way, 
\begin{align}
I_{D}^{r\left(LF\right)} & =g_{0}\left(q_{\perp}\right)\,,\\
I_{G}^{r\left(LF\right)} & =\tilde{g}_{2}\left(q_{\perp}\right)
\,,\nonumber 
\end{align}
are obtained.


\begin{thebibliography}{90}

%\cite{Bartalini:2017jkk}
\bibitem{Bartalini:2017jkk}
  P.~Bartalini and J.~R.~Gaunt,
  %``Multiple Parton Interactions at the LHC,''
  Adv.\ Ser.\ Direct.\ High Energy Phys.\  {\bf 29} (2018) pp.1.
  doi:10.1142/10646
  %%CITATION = doi:10.1142/10646;%%
  %2 citations counted in INSPIRE as of 02 Jan 2019

%\cite{Aad:2013bjm}
\bibitem{Aad:2013bjm}
  G.~Aad {\it et al.} [ATLAS Collaboration],
  %``Measurement of hard double-parton interactions in $W(\to l\nu)$+ 2 jet
  %events at $\sqrt{s}$=7 TeV with the ATLAS detector,''
  New J.\ Phys.\  {\bf 15} (2013) 033038.
%  doi:10.1088/1367-2630/15/3/033038
%  [arXiv:1301.6872 [hep-ex]].
  %%CITATION = doi:10.1088/1367-2630/15/3/033038;%%
  %208 citations counted in INSPIRE as of 26 Mar 2018
  
  %\cite{Paver:1982yp}
\bibitem{Paver:1982yp}
  N.~Paver and D.~Treleani,
  %``Multi - Quark Scattering and Large $p_T$ Jet Production in Hadronic Collisions,''
  Nuovo Cim.\ A {\bf 70} (1982) 215.
%  doi:10.1007/BF02814035
  %%CITATION = doi:10.1007/BF02814035;%%
  %203 citations counted in INSPIRE as of 26 Mar 2018
  
    %\cite{Diehl:2011yj}
%\bibitem{Diehl:2011yj} 
\bibitem{Diehl1}
M.~Diehl, D.~Ostermeier and A.~Schafer,
  %``Elements of a theory for multiparton interactions in QCD,''
  JHEP {\bf 03}, 089 (2012).
%  Erratum: [JHEP {\bf 1603}, 001 (2016)]
%  doi:10.1007/JHEP03(2012)089, 10.1007/JHEP03(2016)001
%  [arXiv:1111.0910 [hep-ph]].
  %%CITATION = doi:10.1007/JHEP03(2012)089, 10.1007/JHEP03(2016)001;%%
  %146 citations counted in INSPIRE as of 20 Jul 2017

%\cite{Guidal:2013rya}
\bibitem{Guidal:2013rya}
  M.~Guidal, H.~Moutarde and M.~Vanderhaeghen,
  %``Generalized Parton Distributions in the valence region from Deeply
  %Virtual Compton Scattering,''
  Rept.\ Prog.\ Phys.\  {\bf 76} (2013) 066202.
%  doi:10.1088/0034-4885/76/6/066202
%  [arXiv:1303.6600 [hep-ph]].
  %%CITATION = doi:10.1088/0034-4885/76/6/066202;%%
  %96 citations counted in INSPIRE as of 26 Mar 2018
  
  
 %\cite{Dupre:2016mai}
\bibitem{Dupre:2016mai}
  R.~Dupr\'e, M.~Guidal and M.~Vanderhaeghen,
  %``Tomographic image of the proton,''
  Phys.\ Rev.\ D {\bf 95} (2017) no.1,  011501.
%  doi:10.1103/PhysRevD.95.011501
%  [arXiv:1606.07821 [hep-ph]].
  %%CITATION = doi:10.1103/PhysRevD.95.011501;%%
  %7 citations counted in INSPIRE as of 26 Mar 2018
      %\cite{Blok:2011bu}

\bibitem{blok1}
%\cite{Blok:2011bu}
%\bibitem{Blok:2011bu} 
  B.~Blok, Y.~Dokshitser, L.~Frankfurt and M.~Strikman,
  %``pQCD physics of multiparton interactions,''
  Eur.\ Phys.\ J.\ C {\bf 72}, 1963 (2012).
%  doi:10.1140/epjc/s10052-012-1963-8
%  [arXiv:1106.5533 [hep-ph]].
  %%CITATION = doi:10.1140/epjc/s10052-012-1963-8;%%
  %81 citations counted in INSPIRE as of 04 Aug 2016

\bibitem{blok2}
%\cite{Blok:2013bpa}
%\bibitem{Blok:2013bpa} 
  B.~Blok, Y.~Dokshitzer, L.~Frankfurt and M.~Strikman,
  %``Perturbative QCD correlations in multi-parton collisions,''
  Eur.\ Phys.\ J.\ C {\bf 74}, 2926 (2014).
%  doi:10.1140/epjc/s10052-014-2926-z
%  [arXiv:1306.3763 [hep-ph]].
  %%CITATION = doi:10.1140/epjc/s10052-014-2926-z;%%
  %52 citations counted in INSPIRE as of 04 Aug 2016

 \bibitem{fabbro}
 %\cite{DelFabbro:2000ds}
%\bibitem{DelFabbro:2000ds} 
  A.~Del Fabbro and D.~Treleani,
  %``Scale factor in double parton collisions and parton densities in transverse 
%space,''
  Phys.\ Rev.\ D {\bf 63}, 057901 (2001).
%  doi:10.1103/PhysRevD.63.057901
%  [hep-ph/0005273].
  %%CITATION = doi:10.1103/PhysRevD.63.057901;%%
  %40 citations counted in INSPIRE as of 20 Jun 2018
  
  
\bibitem{ffe}
%\cite{Rinaldi:2018slz}
%\bibitem{Rinaldi:2018slz} 
  M.~Rinaldi and F.~A.~Ceccopieri,
  %``Hadronic structure from double parton scattering,''
  Phys.\ Rev.\ D {\bf 97}, no. 7, 071501 (2018);
%  doi:10.1103/PhysRevD.97.071501
%  [arXiv:1801.04760 [hep-ph]].
  %%CITATION = doi:10.1103/PhysRevD.97.071501;%%  
  %\cite{Rinaldi:2018bsf}
%\bibitem{Rinaldi:2018bsf}
  M.~Rinaldi and F.~A.~Ceccopieri,
  %``Double parton scattering and the proton transverse structure at the LHC,''
  JHEP {\bf 1909} (2019) 097.
%  [arXiv:1812.04286 [hep-ph]].
  
%\cite{Kasemets:2017vyh}
\bibitem{Kasemets:2017vyh}
  T.~Kasemets and S.~Scopetta,
  %``Parton correlations in double parton scattering,''
  Adv.\ Ser.\ Direct.\ High Energy Phys.\  {\bf 29} (2018) 49
  doi:10.1142/9789813227767\_0004
  [arXiv:1712.02884 [hep-ph]].
  %%CITATION = doi:10.1142/9789813227767_0004;%%
  %6 citations counted in INSPIRE as of 02 Jan 2019
  
  
 
%\cite{Chang:2012nw}
\bibitem{bag}
  H.~M.~Chang, A.~V.~Manohar and W.~J.~Waalewijn,
  %``Double Parton Correlations in the Bag Model,''
  Phys.\ Rev.\ D {\bf 87} (2013) no.3,  034009.
%  doi:10.1103/PhysRevD.87.034009
%  [arXiv:1211.3132 [hep-ph]].
  %%CITATION = doi:10.1103/PhysRevD.87.034009;%%
  %40 citations counted in INSPIRE as of 04 Jun 2018

\bibitem{noi2}
  %\cite{Rinaldi:2013vpa}
%\bibitem{Rinaldi:2013vpa} 
  M.~Rinaldi, S.~Scopetta and V.~Vento,
  %``Double parton correlations in constituent quark models,''
  Phys.\ Rev.\ D {\bf 87}, 114021 (2013).
%  doi:10.1103/PhysRevD.87.114021
%  [arXiv:1302.6462 [hep-ph]].
  %%CITATION = doi:10.1103/PhysRevD.87.114021;%%
  %32 citations counted in INSPIRE as of 27 Jul 2017



 \bibitem{noi1} 
 %\cite{Rinaldi:2014ddl}
%\bibitem{Rinaldi:2014ddl} 
  M.~Rinaldi, S.~Scopetta, M.~Traini and V.~Vento,
  %``Double parton correlations and constituent quark models: a Light Front 
%approach to the valence sector,''
  JHEP {\bf 12}, 028 (2014).
%  doi:10.1007/JHEP12(2014)028
%  [arXiv:1409.1500 [hep-ph]].
  %%CITATION = doi:10.1007/JHEP12(2014)028;%%
  %21 citations counted in INSPIRE as of 21 Jul 2017


\bibitem{kase}
%\bibitem{Kasemets:2016nio} 
  T.~Kasemets and A.~Mukherjee,
  %``Quark-gluon double parton distributions in the light-front dressed quark 
%model,''
  Phys.\ Rev.\ D {\bf 94}, no. 7, 074029 (2016).
%  doi:10.1103/PhysRevD.94.074029
%  [arXiv:1606.05686 [hep-ph]].
  %%CITATION = doi:10.1103/PhysRevD.94.074029;%%
  %4 citations counted in INSPIRE as of 26 Jan 2018
 
   \bibitem{plb} 
  %\cite{Rinaldi:2015cya}
%\bibitem{Rinaldi:2015cya} 
  M.~Rinaldi, S.~Scopetta, M.~Traini and V.~Vento,
  %``Double parton scattering: a study of the effective cross section within a 
%Light-Front quark model,''
  Phys.\ Lett.\ B {\bf 752}, 40 (2016).
%  doi:10.1016/j.physletb.2015.11.031
%  [arXiv:1506.05742 [hep-ph]].
  %%CITATION = doi:10.1016/j.physletb.2015.11.031;%%
  %6 citations counted in INSPIRE as of 07 Feb 2017
  
   \bibitem{JHEP2016} 
 %\cite{Rinaldi:2016jvu}
%\bibitem{Rinaldi:2016jvu} 
  M.~Rinaldi, S.~Scopetta, M.~C.~Traini and V.~Vento,
  %``Correlations in Double Parton Distributions: Perturbative and 
%Non-Perturbative effects,''
  JHEP {\bf 16}, 063 (2016).
%  doi:10.1007/JHEP10(2016)063
%  [arXiv:1608.02521 [hep-ph]].
  %%CITATION = doi:10.1007/JHEP10(2016)063;%%
  %11 citations counted in INSPIRE as of 26 Mar 2018

%\cite{Traini:2016jru}
\bibitem{Traini:2016jru}
  M.~Traini, M.~Rinaldi, S.~Scopetta and V.~Vento,
  %``The effective cross section for double parton scattering 
  %within a holographic AdS/QCD approach,''
  Phys.\ Lett.\ B {\bf 768} (2017) 270.
%  doi:10.1016/j.physletb.2017.02.061
%  [arXiv:1609.07242 [hep-ph]].
  %%CITATION = doi:10.1016/j.physletb.2017.02.061;%%
  %11 citations counted in INSPIRE as of 02 Apr 2018


%\cite{Kirschner:1979im}
\bibitem{Kirschner:1979im}
  R.~Kirschner,
  %``Generalized {Lipatov-Altarelli-Parisi} Equations and Jet Calculus Rules,''
  Phys.\ Lett.\  {\bf 84B} (1979) 266.
%  doi:10.1016/0370-2693(79)90300-9
  %%CITATION = doi:10.1016/0370-2693(79)90300-9;%%
  %78 citations counted in INSPIRE as of 26 Mar 2018

%\cite{Shelest:1982dg}
\bibitem{Shelest:1982dg}
  V.~P.~Shelest, A.~M.~Snigirev and G.~M.~Zinovev,
  %``The Multiparton Distribution Equations in {QCD},''
  Phys.\ Lett.\  {\bf 113B} (1982) 325.
%  doi:10.1016/0370-2693(82)90049-1
  %%CITATION = doi:10.1016/0370-2693(82)90049-1;%%
  %85 citations counted in INSPIRE as of 26 Mar 2018


%\cite{Diehl:2017wew}
\bibitem{Diehl:2017wew}
  M.~Diehl and J.~R.~Gaunt,
  %``Double parton scattering theory overview,''
  Adv.\ Ser.\ Direct.\ High Energy Phys.\  {\bf 29} (2018) 7
 [arXiv:1710.04408 [hep-ph]].
  %%CITATION = doi:10.1142/9789813227767_0002;%%
  %19 citations counted in INSPIRE as of 18 Jun 2019

{
%\cite{Zimmermann:2017ctb}
\bibitem{Zimmermann:2017ctb} 
  C.~Zimmermann [RQCD Collaboration],
  %``Double Parton Distributions of the Pion,''
  PoS LATTICE {\bf 2016}, 152 (2016).
%  doi:10.22323/1.256.0152
%  [arXiv:1701.05479 [hep-lat]].
  %%CITATION = doi:10.22323/1.256.0152;%%
}

%\cite{Bali:2018nde}
\bibitem{Bali:2018nde}
  G.~S.~Bali {\it et al.},
  %``Two-current correlations in the pion on the lattice,''
  JHEP {\bf 1812} (2018) 061
  doi:10.1007/JHEP12(2018)061
  [arXiv:1807.03073 [hep-lat]].
  %%CITATION = doi:10.1007/JHEP12(2018)061;%%
  %2 citations counted in INSPIRE as of 02 Jan 2019

%\cite{Rinaldi:2018zng}
\bibitem{Rinaldi:2018zng}
  M.~Rinaldi, S.~Scopetta, M.~Traini and V.~Vento,
  %``A model calculation of double parton distribution functions of the pion,''
  Eur.\ Phys.\ J.\ C {\bf 78} (2018) no.9,  781.
%  doi:10.1140/epjc/s10052-018-6256-4
%  [arXiv:1806.10112 [hep-ph]].
  %%CITATION = doi:10.1140/epjc/s10052-018-6256-4;%%
  %3 citations counted in INSPIRE as of 02 Jan 2019
  
%\cite{Klevansky:1992qe}
\bibitem{Klevansky:1992qe}
  S.~P.~Klevansky,
  %``The Nambu-Jona-Lasinio model of quantum chromodynamics,''
  Rev.\ Mod.\ Phys.\  {\bf 64} (1992) 649.
 % doi:10.1103/RevModPhys.64.649
  %%CITATION = doi:10.1103/RevModPhys.64.649;%%
  %1414 citations counted in INSPIRE as of 17 May 2018  
 

 
 \bibitem {Davidson:2001cc} 
R.~M.~Davidson and E.~Ruiz Arriola,
  %``Parton distributions functions of pion, kaon and eta pseudoscalar mesons in the NJL model,''
  Acta Phys.\ Polon.\ B {\bf 33} (2002) 1791.
%  [hep-ph/0110291].
  %%CITATION = HEP-PH/0110291;%%

\bibitem {Theussl:2002xp}
L. Theussl, S. Noguera and V. Vento,
%  \emph{Generalized parton distributions of the pion in a Bethe-Salpeter approach},
 Eur. Phys. J. A \textbf{20} (2004) 483. 
%  doi:10.1140/epja/i2003-10174-3
%  [arXiv:nucl-th/0211036].
%%CITATION = EPHJA,A20,483;%%

\bibitem {RuizArriola:2002bp} 
E.~Ruiz Arriola and W.~Broniowski,
 % \emph{Pion light-cone wave function and pion distribution amplitude in the Nambu-Jona-Lasinio model},
 Phys.\ Rev.\ D \textbf{66} (2002) 094016. 
 %  doi:10.1103/PhysRevD.66.094016
 % [arXiv:hep-ph/0207266].
%%CITATION = PHRVA,D66,094016;%%

%\cite{Courtoy:2008nf}
\bibitem{Courtoy:2008nf}
  A.~Courtoy and S.~Noguera,
 % \emph{Enhancement effects in exclusive pi pi and rho pi production in gamma* gamma scattering},
  Phys.\ Lett.\ B {\bf 675} (2009) 38.
 % doi:10.1016/j.physletb.2009.03.070
 % [arXiv:0811.0550 [hep-ph]].
  %%CITATION = doi:10.1016/j.physletb.2009.03.070;%%
  %7 citations counted in INSPIRE as of 30 Nov 2017

%\cite{Courtoy:2007vy}
\bibitem{Courtoy:2007vy}
A.~Courtoy and S.~Noguera,
  %``The Pion-photon transition distribution amplitudes in the Nambu-Jona Lasinio model,''
Phys.\ Rev.\ D {\bf 76} (2007) 094026.
%  doi:10.1103/PhysRevD.76.094026
%  [arXiv:0707.3366 [hep-ph]].
  %%CITATION = doi:10.1103/PhysRevD.76.094026;%%
  %41 citations counted in INSPIRE as of 08 Mar 2019
  
\bibitem{Courtoythesis} 
A.~Courtoy, Ph. D. Thesis, Valencia University, 2009. 
\\
http://arxiv.org/abs/arXiv:1010.2974

%\cite{Noguera:2011fv}
\bibitem{Noguera:2011fv} 
 S.~Noguera and S.~Scopetta,
  %``The eta-photon transition form factor,''
  Phys.\ Rev.\ D {\bf 85} (2012) 054004.
%  doi:10.1103/PhysRevD.85.054004
%  [arXiv:1110.6402 [hep-ph]].
  %%CITATION = doi:10.1103/PhysRevD.85.054004;%%
  
  %\cite{Weigel:1999pc}
\bibitem{Weigel:1999pc}
  H.~Weigel, E.~Ruiz Arriola and L.~P.~Gamberg,
 % \emph{Hadron structure functions in a chiral quark model: Regularization, scaling and sum rules},
  Nucl.\ Phys.\ B{\bf 560} (1999) 383.
%  doi:10.1016/S0550-3213(99)00426-5
%  [hep-ph/9905329].
  %%CITATION = doi:10.1016/S0550-3213(99)00426-5;%%
  %56 citations counted in INSPIRE as of 29 Jan 2018
  
   \bibitem{ns} 
%\cite{Noguera:2015iia}
  S.~Noguera and S.~Scopetta,
 % \emph{Pion transverse momentum dependent parton distributions in the Nambu and Jona-Lasinio model},
 JHEP {\bf 1511} (2015) 102.
 % doi:10.1007/JHEP11(2015)102
 % [arXiv:1508.01061 [hep-ph]].
  
  %\cite{Broniowski:2017gfp}
\bibitem{Broniowski:2017gfp}
  W.~Broniowski and E.~Ruiz Arriola,
  %``Partonic quasidistributions of the proton and pion from transverse-momentum distributions,''
  Phys.\ Rev.\ D {\bf 97} (2018) no.3,  034031.
 % doi:10.1103/PhysRevD.97.034031
 % [arXiv:1711.03377 [hep-ph]].
  %%CITATION = doi:10.1103/PhysRevD.97.034031;%%
  %4 citations counted in INSPIRE as of 29 Jan 2018

%\cite{Ceccopieri:2018nop}
\bibitem{Ceccopieri:2018nop} 
  F.~A.~Ceccopieri, A.~Courtoy, S.~Noguera and S.~Scopetta,
  %``Pion nucleus Drell–Yan process and parton transverse momentum in the pion''.
  Eur.\ Phys.\ J.\ C {\bf 78}, no. 8, 644 (2018).
%  doi:10.1140/epjc/s10052-018-6115-3
%  [arXiv:1801.07682 [hep-ph]].
  %%CITATION = doi:10.1140/epjc/s10052-018-6115-3;%%
  %2 citations counted in INSPIRE as of 08 Jan 2019

\bibitem{gaunt}
%\cite{Gaunt:2009re}
  J.~R.~Gaunt and W.~J.~Stirling,
  %``Double Parton Distributions Incorporating Perturbative QCD Evolution and Momentum and Quark Number Sum Rules,''
  JHEP {\bf 1003} (2010) 005.
%  doi:10.1007/JHEP03(2010)005
%  [arXiv:0910.4347 [hep-ph]].
  %%CITATION = doi:10.1007/JHEP03(2010)005;%%
  %161 citations counted in INSPIRE as of 25 Jun 2019


\bibitem{Radyushkin:1997ki}A.~V.~Radyushkin, 
%``Nonforward parton
%distributions,'' 
Phys. Rev. D \textbf{56} (1997) 5524. %doi:10.1103/PhysRevD.56.5524
%{[}hep-ph/9704207{]}. %%CITATION = doi:10.1103/PhysRevD.56.5524;%%
%1054 citations counted in INSPIRE as of 03 Feb 2019

\bibitem{Ji:1998pc}X.~D.~Ji, 
%``Off forward parton distributions,''
J. Phys. G \textbf{24} (1998) 1181.
%doi:10.1088/0954-3899/24/7/002
%{[}hep-ph/9807358{]}. %%CITATION = doi:10.1088/0954-3899/24/7/002;%%
%448 citations counted in INSPIRE as of 03 Feb 2019



%\cite{Diehl:2013mla}
\bibitem{Diehl:2013mla} 
  M.~Diehl and T.~Kasemets,
  %``Positivity bounds on double parton distributions,''
  JHEP {\bf 1305}, 150 (2013)
  doi:10.1007/JHEP05(2013)150
  [arXiv:1303.0842 [hep-ph]].
  %%CITATION = doi:10.1007/JHEP05(2013)150;%%
  %26 citations counted in INSPIRE as of 06 Mar 2019

\bibitem{inprog}
A. Courtoy, S. Noguera and S. Scopetta, in progress.

\bibitem{private} 
C. Zimmermann, private communication.


%\cite{Burkardt:1995ct}
\bibitem{Burkardt:1995ct}
  M.~Burkardt,
  %``Light front quantization,''
  Adv.\ Nucl.\ Phys.\  {\bf 23} (1996) 1 % doi:10.1007/0-306-47067-5_1
  [hep-ph/9505259].
  %%CITATION = doi:10.1007/0-306-47067-5_1;%%
  %185 citations counted in INSPIRE as of 04 Aug 2019

%\cite{Burkardt:1997bd}
\bibitem{Burkardt:1997bd}
  M.~Burkardt,
  %``Much ado about nothing: Vacuum and renormalization on the light front,''
  In *Seoul 1997, QCD, lightcone physics and hadron phenomenology* 170-199
  [hep-ph/9709421].
  %%CITATION = HEP-PH/9709421;%%
  %15 citations counted in INSPIRE as of 04 Aug 2019
  
{  
%\cite{Itakura:2000te}
\bibitem{Itakura:2000te} 
K.~Itakura and S.~Maedan,
  %``Dynamical chiral symmetry breaking on the light front. 2. The Nambu-Jona-Lasinio model,''
Phys.\ Rev.\ D {\bf 62}, 105016 (2000).
%doi:10.1103/PhysRevD.62.105016
%  [hep-ph/0004081].
  %%CITATION = doi:10.1103/PhysRevD.62.105016;%%
  %10 citations counted in INSPIRE as of 06 Mar 2019
}

\end{thebibliography}
\end{document}